\documentclass[reprint,prd,nofootinbib,superscriptaddress]{revtex4}
\usepackage[utf8]{inputenc}
\usepackage{eurosym}
\usepackage{amsfonts}
\usepackage{amsmath}
\usepackage{amssymb}
\usepackage[english]{babel}
\usepackage{graphicx}
\usepackage{epsfig}
\usepackage{bm}
\usepackage{verbatim}
\usepackage{booktabs}
\usepackage{multirow}
\usepackage{xcolor}
\usepackage[colorlinks=true,urlcolor=red,citecolor=red]{hyperref}
\usepackage[font=small]{caption}
\usepackage{float}
\usepackage{blindtext}
\usepackage{booktabs}
\usepackage{subfigure}
\usepackage{tikz}
\usepackage{pgffor}

\usetikzlibrary{arrows,shapes}
\usetikzlibrary{trees}
\usetikzlibrary{matrix,arrows} 
\usetikzlibrary{positioning}
\usetikzlibrary{calc,through}
\usetikzlibrary{decorations.pathreplacing}  
\usetikzlibrary{decorations.pathmorphing}
\usetikzlibrary{decorations.markings}
\tikzset{
    vector/.style={decorate, decoration={snake}, draw},
provector/.style={decorate, decoration={snake,amplitude=2.5pt}, draw},
antivector/.style={decorate, decoration={snake,amplitude=-2.5pt}, draw},
    fermion/.style={draw=black, postaction={decorate},
        decoration={markings,mark=at position .55 with {\arrow[draw=black]{>}}}},
    fermionbar/.style={draw=black, postaction={decorate},
        decoration={markings,mark=at position .55 with {\arrow[draw=black]{<}}}},
    fermionnoarrow/.style={draw=black},
    gluon/.style={decorate, draw=black,
        decoration={coil,amplitude=4pt, segment length=5pt}},
    scalar/.style={dashed,draw=black, postaction={decorate},
        decoration={markings,mark=at position .55 with {\arrow[draw=black]{>}}}},
    scalarbar/.style={dashed,draw=black, postaction={decorate},
        decoration={markings,mark=at position .55 with {\arrow[draw=black]{<}}}},
    scalarnoarrow/.style={dashed,draw=black},
    electron/.style={draw=black, postaction={decorate},
        decoration={markings,mark=at position .55 with {\arrow[draw=black]{>}}}},
bigvector/.style={decorate, decoration={snake,amplitude=4pt}, draw},
}\usetikzlibrary{decorations.markings}
\tikzstyle{block} = [draw, rectangle, 
    minimum height=3em, minimum width=6em]

\newcommand{\bc}{\begin{center}}
\newcommand{\ec}{\end{center}}

\newcommand{\Om}{\Omega}

\newcommand{\mathsym}[1]{}

\newcommand{\bea}{\begin{eqnarray}}
\newcommand{\eea}{\end{eqnarray}}
\input{tcilatex}
\begin{document}

\title{Extended IDM theory with low scale seesaw mechanisms}
\author{D. T. Huong}
\email{dthuong@iop.vast.vn}
\affiliation{Institute of Physics, VAST, 10 Dao Tan, Giang Vo, Hanoi, Vietnam}
\date{\today }
\author{A.E. C\'{a}rcamo Hern\'{a}ndez}
\email{antonio.carcamo@usm.cl}
\affiliation{Universidad T\'{e}cnica Federico Santa Mar\'{\i}a, Casilla 110-V, Valpara%
\'{\i}so, Chile}
\affiliation{Centro Cient\'ifico-Tecnol\'ogico de Valpara\'iso, Casilla 110-V, Valpara%
\'{\i}so, Chile}
\affiliation{Millennium Institute for Subatomic Physics at high energy frontier - SAPHIR,
Fernandez Concha 700, Santiago, Chile}
\author{H. T. Hung}
\email{hathanhhung@hpu2.edu.vn}
\affiliation{Department of Physics, Hanoi Pedagogical University 2, Xuan Hoa, Phu Tho, Vietnam}
\author{T. T. Hieu}
\email{tranhieusp2@gmail.com}
\affiliation{Graduate University of Science and Technology, Vietnam Academy of Science
and Technology, 18 Hoang Quoc Viet, Cau Giay, Hanoi, Vietnam}
\affiliation{Department of Physics, Hanoi Pedagogical University 2, Xuan Hoa, Phu Tho, Vietnam}
\author{Nicol\'{a}s A. P\'{e}rez-Julve}
\email{nicolasperezjulve@gmail.com}
\affiliation{Universidad T\'{e}cnica Federico Santa Mar\'{\i}a, Casilla 110-V, Valpara%
\'{\i}so, Chile}
\affiliation{Millennium Institute for Subatomic Physics at high energy frontier - SAPHIR,
Fernandez Concha 700, Santiago, Chile}
\author{N. T. Duy}
\email{ntduy@iop.vast.vn}
\affiliation{Institute of Physics, VAST, 10 Dao Tan, Ba Dinh, Hanoi, Vietnam}

\begin{abstract}
We have developed an extension of the inert doublet model in which the
CP-phases in the weak sector are generated from one-loop level corrections
mediated by dark fields, while the strong-CP phase remain vanishing at three-loop. In
this framework, the tiny masses of the active neutrinos are produced through
a radiative inverse seesaw mechanism at a two-loop level, the masses of the
first and second families of SM-charged fermions arise from a one-loop level
radiative seesaw mechanism, and the third generation of SM charged fermion
masses are generated at tree level. We have demonstrated that the proposed
model successfully accounts for SM fermion masses and mixings. The radiative
nature of the seesaw mechanisms is attributed to preserved discrete
symmetries, which are required for ensuring the stability of fermionic and
scalar dark matter candidates. The preserved discrete symmetries also allow
for multi-component dark matter, whose annihilation processes permits to
successfully reproduce the measured amount of dark matter relic abundance
for an appropriate region of parameter space, which has shown to be
compatible with current dark matter direct detection limits. Besides that,
we explore the model's ability to explain the $95$ GeV diphoton excess
observed by the CMS collaboration, showing that it readily accommodates this
anomaly. We have shown that charged lepton flavor violating decays acquire
rates within the current experimental sensitivity.
\end{abstract}

\maketitle

\section{\label{intro}Introduction}

Despite the great success of the Standard Model (SM) as a theory of strong
and electroweak interactions whose predictions have been experimentally
verified with the highest degree of accuracy, there are several issues that
the SM is unable to explain, such as the smallness of neutrino masses ~\cite%
{McDonald:2016ixn}, the current amount of the dark matter relic density ~%
\cite{Bertone:2004pz}, the tiny values of the electric dipole moments of the
neutron and the elementary particles \cite{Pendlebury:2015lrz}, and the
SM-charged fermion mass and mixing hierarchy. In the SM, the gauge
invariance does not restrict the flavor structure of the Yukawa
interactions. In particular, the Yukawa couplings of fermions to the Higgs
field exhibit a wide range of values without any apparent underlying
principle. While the discovery of the Higgs boson \cite{ATLAS:2012yve,
CMS:2012qbp} confirms the existence of these interactions, current collider
experiments are primarily sensitive to the Yukawa couplings of
third-generation fermions \cite{CMS:2018uxb, ATLAS:2015xst, CMS:2018uag,
PhysRevLett.10.531, CMS:2017zyp, ATLAS:2018mme, ATLAS:2018kot}. The
hypothesis that a more fundamental theory generates SM Yukawa couplings
offers a promising avenue for understanding the flavor puzzle \cite%
{Balakrishna:1988ks, Ma:1988fp, Kitabayashi:2000nq, Chang:2006aa,
CarcamoHernandez:2013zrj, CarcamoHernandez:2013yiy, Campos:2014lla,
Boucenna:2014ela, Okada:2015bxa,Wang:2015saa,
Arbelaez:2016mhg,Nomura:2016emz,Kownacki:2016hpm, Nomura:2016ezz,
CarcamoHernandez:2020ehn, Hernandez:2021zje, CarcamoHernandez:2021tlv,
CarcamoHernandez:2023atk, CarcamoHernandez:2024edi, CarcamoHernandez:2024vcr}%
. Such fundamental theory is expected to be 
highly symmetric 
at high energies, these Yukawa terms either vanish or converge to a common
value. Theories 
focusing on explaining the origin of the SM fermion structure have been
proposed and analyzed in detail 
in \cite{CarcamoHernandez:2015hjp, Camargo-Molina:2016yqm,
Camargo-Molina:2016bwm, CarcamoHernandez:2017kra, Dev:2018pjn,
CarcamoHernandez:2018hst, CarcamoHernandez:2016pdu,
CarcamoHernandez:2017owh, CarcamoHernandez:2017cwi,
CarcamoHernandez:2019cbd, Arbelaez:2019ofg, Hernandez:2021iss,
Hernandez:2021xet, Binh:2024lez, CarcamoHernandez:2020owa,
CarcamoHernandez:2019lhv, Hernandez:2021tii, Hernandez:2021uxx,
CarcamoHernandez:2022vjk, CarcamoHernandez:2023oeq,
CarcamoHernandez:2023wzf, C:2024exl}. These models generate the top quark
mass at the tree level. In contrast, the masses of the remaining particles,
such as the bottom quark, tau lepton, and muon, arise at one-loop or
higher-order levels, leading to a hierarchical mass spectrum.


The origin of quark mixing and the size of CP violation in this sector are
also related issues. Non-perturbative effects in QCD can result in $\text{P}$%
- and $\text{CP}$-violations, indicated by a parameter $\theta$, which is
the sum of two terms: $\theta_{\text{QCD}}$ and $\theta_{\text{QED}}$. The $%
\theta_{\text{QED}}$ appears due to the chiral rotation of the quark fields,
which render the real and positive quark masses. The $\theta_{\text{QCD}}$
is the value of $\text{P}$- and $\text{CP}$- violating angles of QCD vacuum.
The appearance of $\theta$ induces the neutron electric dipole moment
(nEDM), which is constrained by the experiment \cite{Pendlebury:2015lrz} as $%
d_n \lesssim 3 \times 10^{-26}e$. It implies that $|\theta| \lesssim
10^{-10} $, which appears unnatural, and then corresponds to the strong CP
problem. The popular solution to the strong CP problem in QCD is an
assumption about the existence of global Peccei-Quinn (PQ) symmetry \cite%
{Peccei:1977ur,Peccei:1977hh,Weinberg:1977ma} 
or spontaneous CP symmetry breaking, which requires to extend the SM
particle content by adding a complex scalar field responsible for such CP
breaking \cite{Nelson:1983zb, Barr:1984qx, Nelson:1984hg, Barr:1984fh}. 
Recent studies performed in \cite{Camara:2023hhn} have proposed a new
solution to the strong CP problem based on the existence of a dark sector.
Both the CP symmetry and $Z_8$ are spontaneously broken in such a way to
give rise to a residual $Z_2$ discrete symmetry that allows the stability of
the 
dark matter. The strong CP phase can arise via loop corrections mediated by
dark fields. In the following work, we show that in our scenario, at the
tree level, the active sector conserves the CP symmetry, then implying that
the 
parameters $\theta_{\text{ QCD}}$ and $\theta_{\text{ QED}}$ vanish at tree
level. The explicit CP violation appears in the dark sector. The CP
violation can be transmitted to the SM quark sector via loop corrections
mediated by the dark fields.

We achieve this in an extended inert doublet model (IDM) framework where the
CP phases in the weak sector arise at one-loop level, whereas the strong CP
phase remain vanishing at the three-loop level. In that theory, the tiny active neutrino
masses are originated from an inverse seesaw mechanism at two-loop level;
the first and second generation of SM-charged fermion masses arise at one
loop, whereas the third family of SM-charged fermions obtains tree level
masses. 
It is worth mentioning that in most of the extended IDM, the tiny active
neutrino masses are produced by a radiative seesaw mechanism at one loop 
\cite{Balakrishna:1988ks, Ma:1988fp, Ma:1989ys, Ma:1990ce, Ma:1998dn,
Tao:1996vb, Ma:2006km, Gu:2007ug, Ma:2008cu, Hirsch:2013ola, Aranda:2015xoa,
Restrepo:2015ura, Longas:2015sxk, Fraser:2015zed, Fraser:2015mhb,
Wang:2015saa, Arbelaez:2016mhg, vonderPahlen:2016cbw, Nomura:2016emz,
Kownacki:2016hpm, Nomura:2017emk, Nomura:2017vzp, Bernal:2017xat,
Wang:2017mcy, Bonilla:2018ynb, Calle:2018ovc, Avila:2019hhv,
CarcamoHernandez:2018aon, Alvarado:2021fbw, Arbelaez:2022ejo,
Cepedello:2022xgb, CarcamoHernandez:2022vjk, Leite:2023gzl}, then requiring
very small neutrino Yukawa couplings (of the order of the electron Yukawa
coupling) or a tiny mass difference between the dark scalars and pseudosalar
seesaw messengers, to successfully accommodate the experimental values of
the neutrino mass squared splittings. Two-loop neutrino mass models have
been proposed and analyzed in the literature~\cite{Bonilla:2016diq,
Baek:2017qos, Saad:2019vjo, Nomura:2019yft, Arbelaez:2019wyz, Saad:2020ihm,
Xing:2020ezi, Chen:2020ptg, Nomura:2020dzw} in order to provide a more
natural explanation for the tiny values of the active neutrino masses than
those relying on radiative seesaw mechanisms at one loop level. In this
work, we consider an extension of the IDM where the inclusion of gauge
singlet scalars augments the scalar sector, and the fermion sector is
enlarged by charged vector-like fermions and right-handed Majorana
neutrinos. In that theory, the SM gauge symmetry is supplemented by
including the spontaneously broken $U(1)_X$ global symmetry and the
preserved $Z_2$ symmetry. The charged vector-like fermions, dark scalars,
and pseudoscalars mediate a one-loop level radiative seesaw mechanism that
yields the masses of the first and second generation of SM-charged fermions.
Additionally, the charged vector-like fermions and the dark scalars and
pseudoscalars provide radiative corrections to the CP phases in the weak
sector at one level and mantain the strong CP phase vanishing at three loop level. On the other hand, the tiny masses of the active neutrinos are
produced by an inverse seesaw mechanism at two-loop level where the lepton
number violating Majorana mass terms arise at two loops. This is achieved
thanks to the preserved $Z_2$ symmetry as well as to the remnant $%
Z_{2}^{\prime }$ symmetry arising from the spontaneous breaking of the
global $U(1)_X$ symmetry. These preserved discrete symmetries ensure the
radiative nature of the above-mentioned seesaw mechanisms and the stability
of the dark matter candidates. The model is compatible with the current
pattern of SM fermion masses and mixings, with the constraints arising from
charged lepton flavor violation, dark matter relic density, and direct
detection. It also successfully accommodates the $95$ GeV diphoton excess.
This paper is organized as follows. In section \ref{model}, we describe the
model in detail. Its implications in SM fermion masses and mixings are
discussed in section \ref{fermionmassesandmixings}. Section \ref{strongCP}
provides a detailed discussion of how the strong CP problem is addressed in
the model under consideration. The consequences of the model in dark matter,
charged lepton flavor violation, and the $95$ GeV diphoton excess are
discussed in sections \ref{dm}, \ref{sec:diphoton-excess}, and \ref{cLFV},
respectively. We state our conclusions in section \ref{conclusions}.

\section{Model}

\label{model}

\subsection{Particle content}

We start this section by explaining the reasoning that
justifies the inclusion of extra scalar and fermions needed for the
implementation of the radiative seesaw mechanisms that generates one loop
level masses for the first and second families of SM charged fermions,
two-loop level tiny active neutrino masses as well as two-loop level mixing
mass terms between SM quarks and some heavy quarks. Besides that, the masses
of the third generation of the SM charged fermions will be generated at tree
level by the following operators:%
\begin{equation}
\overline{q}_{iL}\widetilde{\phi }u_{3R},\hspace{1cm}\hspace{1cm}\overline{q}%
_{iL}\phi d_{3R},\hspace{1cm}\hspace{1cm}\overline{l}_{iL}\phi l_{3R},%
\hspace{1cm}\hspace{1cm}i=1,2,3.
\end{equation}%
where $\phi $ is the SM $SU\left( 2\right) _{L}$ scalar doublet, which is
expanded as follows 
\begin{equation}
\phi =\left( 
\begin{array}{c}
\phi ^{+} \\ 
\frac{v+\phi _{R}^{0}+i\phi _{I}^{0}}{\sqrt{2}}%
\end{array}%
\right) .
\end{equation}%
whereas $q_{iL}$and $l_{iL}$ are the SM quark and lepton doublets,
respectively. They are defined as follows: 
\begin{equation}
q_{iL}=\left( 
\begin{array}{c}
u_{iL} \\ 
d_{iL}%
\end{array}%
\right) ,\hspace{1cm}\hspace{1cm}l_{iL}=\left( 
\begin{array}{c}
\nu _{iL} \\ 
e_{iL}%
\end{array}%
\right) ,\hspace{1cm}\hspace{1cm}i=1,2,3.
\end{equation}%
To generate one loop level masses for the first and second family of SM
charged fermions, we need to forbid the following operators: 
\begin{equation}
\overline{q}_{iL}\widetilde{\phi }u_{nR},\hspace{1cm}\hspace{1cm}\overline{q}%
_{iL}\phi d_{nR},\hspace{1cm}\hspace{1cm}\overline{l}_{iL}\phi l_{nR},%
\hspace{1cm}\hspace{1cm}n=1,2.
\end{equation}%
at tree level and to allow other operators instead, which are described in
the following. In order to successfully implement the radiative seesaw
mechanism that yields one loop masses for the first and second families of
SM charged fermions, we need the following operators: 
\begin{eqnarray}
&&\overline{q}_{iL}\widetilde{\eta }T_{kR},\hspace{1cm}\hspace{1cm}\overline{%
T}_{kL}\varphi _{2}^{\ast }u_{nR},\hspace{1cm}\hspace{1cm}\overline{T}%
_{kL}\sigma ^{\ast }T_{sR},\hspace{1cm}\hspace{1cm}k,s,n=1,2.  \notag \\
&&\overline{q}_{iL}\eta B_{kR},\hspace{1cm}\hspace{1cm}\overline{B}%
_{kL}\varphi _{2}d_{nR},\hspace{1cm}\hspace{1cm}\overline{B}_{kL}\sigma
B_{nR},\hspace{1cm}\hspace{1cm}i=1,2,3  \notag \\
&&\overline{l}_{iL}\eta E_{kR},\hspace{1cm}\hspace{1cm}\overline{E}%
_{kL}\varphi _{2}l_{nR},\hspace{1cm}\hspace{1cm}\overline{E}_{kL}\sigma
E_{nR},\hspace{1cm}\hspace{1cm}
\end{eqnarray}%
This implies an extension of the SM gauge symmetry by the inclusion of extra 
$U\left( 1\right) _{X}\times Z_{2}$ symmetry, with $U\left( 1\right) _{X}$
assumed to be global. We further assume that the global $U\left( 1\right)
_{X}$ symmetry spontaneously breaks down to a preserved $Z_{2}^{\prime }$
symmetry, which is needed to ensure the radiative nature of the seesaw
mechanism that generates one loop level masses for the first generation of
SM charged fermions. The supplementary $Z_{2}$ symmetry is crucial for ensuring the two loop level nature of the inverse seesaw
mechanism that yields the active neutrino masses. The successful
implementation of the above mentioned radiative seesaw mechanism requires to
enlarge the SM scalar spectrum by including an inert $SU\left( 2\right) _{L}$
scalar doublet $\eta $ as well as the electrically neutral scalar singlets $%
\varphi _{1}$,\ $\varphi _{2}$,\ $\varphi _{3}$ and $\sigma $. In addition,
the fermion sector of the SM has to be augmented as well by including the
following charged vector like fermions: two up type quarks $T_{k}$; two down
type quarks $B_{k}$ and two exotic charged leptons $E_{k}$ ($k=1,2$) in
singlet representations of $SU\left( 2\right) _{L}$. Besides that,
implementing the two-loop level inverse seesaw mechanism to produce tiny
masses of the light active neutrinos, requires the following structure for
the full neutrino mass matrix in the basis $\left( \nu _{L},\nu
_{R}^{C},N_{R}^{C}\right) $: 
\begin{equation}
M_{\nu }=\left( 
\begin{array}{ccc}
0_{3\times 3} & m_{\nu D} & 0_{3\times 3} \\ 
m_{\nu D}^{T} & 0_{2\times 2} & M \\ 
0_{3\times 3} & M^{T} & \mu%
\end{array}%
\right) ,  \label{Mnufull}
\end{equation}%
where $\nu _{iL}$ ($i=1,2,3$) correspond to the active neutrinos, whereas $%
\nu _{iR}$ and $N_{iR}$ ($i=1,2$) are the sterile neutrinos. Furthermore,
the entries of the full neutrino mass matrix of Eq. (\ref{Mnufull}) should
obey the hierarchy $\mu _{nk}<<\left( m_{\nu D}\right) _{in}<<M_{nk}$ ($%
i=1,2,3$, $n,k=1,2$), where the submatrices $m_{\nu D}$ and $M$ are
generated at tree level, whereas the Majorana mass submatrix $\mu $
associated with the breaking of the lepton number in two units is
radiatively generated at two loop level. Consequently in order to
successfully implement the above described inverse seesaw mechanism, we need
the following operators: 
\begin{eqnarray}
&&\overline{l}_{iL}\widetilde{\phi }\nu _{nR},\hspace{1cm}\hspace{1cm}\nu
_{nR}\sigma ^{\ast }\overline{N_{kR}^{C}},\hspace{1cm}\hspace{1cm}\overline{N%
}_{nR}\Psi _{kR}^{C}\varphi _{1},\hspace{1cm}\hspace{1cm}k,n=1,2.  \notag \\
&&\overline{\Psi }_{nR}\varphi _{3}\Omega _{kR}^{C},\hspace{1cm}\hspace{1cm}%
\overline{\Omega }_{nR}\sigma \Omega _{kR}^{C},\hspace{1cm}\hspace{1cm}%
i=1,2,3\hspace{1cm}\hspace{1cm}
\end{eqnarray}%
where $\nu _{nR}$, $\Omega _{nR}$, $\Psi _{nR}$ ($n=1,2$)\ are right handed
Majorana neutrinos which are required to be added to the fermionic spectrum
of the SM.

The model under consideration corresponds to an extended inert doublet
model(IDM). Including electrically neutral scalar singlets enlarges the
scalar sector, and the fermion sector is augmented by adding charged vectors
like fermions and right-handed Majorana neutrinos. The SM gauge symmetry is
extended by the inclusion of the spontaneously broken global $U\left(
1\right) _{X}$ symmetry and the preserved $Z_{2}$ discrete symmetry. In the
model under consideration, the first and second families of SM-charged
fermions get one loop level masses, thanks to the conserved $Z_{2}^{\prime }$
symmetry arising from the spontaneous symmetry breaking of $U\left( 1\right)
_{X}$. In contrast, the third generation of SM-charged fermions obtain their
masses at tree level. The masses of the light-active neutrinos arise from a
radiative inverse seesaw mechanism at a two-loop level. Furthermore, in the
considered model, mixing mass terms between SM quarks and some heavy quarks
are generated at the two-loop level and involve complex parameters arising
from a CP violating scalar potential, then allowing to address the strong CP
problem. The scalar, quark, and leptonic spectrum of the model, as well as
their assignments under the $SU\left( 3\right) _{C}\times SU\left( 2\right)
_{L}\times U\left( 1\right) _{Y}\times U\left( 1\right) _{X}\times Z_{2}$
symmetry 
are displayed in Tables \ref{scalars}, \ref{quarks} and \ref{leptons},
respectively. \newline
\space The SM $SU(2)$ scalar doublet $\phi $ and the scalar singlet $\sigma $
develop the following vacuum expectation values (VEVs):
\begin{equation*}
\langle \phi \rangle =\left( 
\begin{array}{c}
0 \\ 
\frac{v}{\sqrt{2}}%
\end{array}%
\right) ,\hspace{1cm}\langle \sigma \rangle =\frac{v_{\sigma }}{\sqrt{2}}
\end{equation*}%
producing the following scheme of symmetry breaking: 
\begin{eqnarray}
SU(3)_{c}\times SU(2)_{L} &\times &U(1)_{Y}\times U(1)_{X}\times Z_{2} 
\notag \\
&\downarrow &v_{\sigma }  \notag \\
SU(3)_{c}\times SU(2)_{L} &\times &U(1)_{Y}\times Z_{2}^{\prime }\times Z_{2}
\\
&\downarrow &v  \notag \\
SU(3)_{c}\times U(1)_{Q} &\times &Z_{2}^{\prime }\times Z_{2}
\end{eqnarray}%
It is assumed that the global $U\left( 1\right) _{X}$ symmetry is
spontaneously broken down to the preserved $Z_{2}^{\prime }$ symmetry by the
vacuum expectation value (VEV) of the scalar singlet $\sigma $. We note that
the electric charge operator, $Q=T_{3}+Y$, combines isospin and hypercharge,
whereas $Z_{2}^{\prime }$ is the residual symmetry resulting from the
spontaneous breaking of $U(1)_{X}$, takes the form $Z_{2}^{\prime }=e^{i\pi
X}=(-1)^{X}$, then implying that the $Z_{2}^{\prime }$ field assignments of
the model are defined as $\left( -1\right) ^{X}$ being $X$ the corresponding 
$U(1)_{X}$ charges. Hence, the particles with even $X-$charge carry a zero $%
Z_{2}^{\prime }-$ charge, while the particles with odd $X-$charge carry a
unit $Z_{2}^{\prime }$ charge. Due to the preserved $Z_{2}^{\prime }\times
Z_{2}$ symmetry neither the extra $SU(2)$ scalar doublet $\eta $, nor the
singlet scalar fields $\varphi _{1}$, $\varphi _{2}$, $\varphi _{3}$ acquire
vacuum expectation values since they carry non-trivial charges under this
preserved $Z_{2}^{\prime }\times Z_{2}$ symmetry. In what follows, we
justify the particle content of the model. The scalar fields $\eta $ and $%
\varphi _{2}$ are needed for the implementation of the one-loop level
radiative seesaw mechanisms that yield the first and second generations of
SM charged fermion masses. Such radiative seesaw mechanisms are also induced
by the charged vector-like fermions $T_{n}$, $B_{n}$ and $E_{n}$ as
indicated in the Feynman diagrams of Figures (\ref{Quark-oneloop}) and (\ref%
{charged-lepton}). Besides that, the scalar singlets $\varphi _{1}$ and $%
\varphi _{3}$, together with the neutral leptons $\Omega _{nR}$ and $\Psi
_{nR}$ ($n=1,2$) mediate the radiative seesaw mechanism at two-loop level,
as indicated in Figure \ref{muterm}, that results in a dynamical generation
of the lepton number violating $\mu _{nk}\overline{N}_{nR}\sigma N_{kR}^{C}$
Majorana mass terms, then yielding an inverse seesaw mechanism at two-loop
level.

Notice that the preserved $Z_{2}^{\prime }\times Z_{2}$ symmetry is also
crucial for avoiding the appearance of tree-level masses for active
neutrinos as well as for the first and second families of SM-charged
fermions. It is worth mentioning that the preserved $Z_{2}$ and $%
Z_{2}^{\prime }$ discrete symmetries are crucial for ensuring the radiative
nature of the one-loop level radiative seesaw mechanisms that yield the
masses of the first and second-generation of SM-charged fermions, as well as
the inverse seesaw mechanism \cite%
{Mohapatra:1986bd,Malinsky:2005bi,Malinsky:2009df,Guo:2012ne,Law:2012mj,Baldes:2013eva,Abada:2014vea,Mandal:2019oth,CarcamoHernandez:2019lhv,Abada:2021yot,Hernandez:2021xet,Hernandez:2021kju,Bonilla:2023egs,Bonilla:2023wok,Abada:2023zbb,Binh:2024lez,Gomez-Izquierdo:2024apr}
that produces the tiny active neutrino masses. Furthermore, because of the
conserved $Z_{2}^{\prime }\times Z_{2}$ symmetry, the model offers a natural
stability mechanism for two-component dark matter in which a dark matter
component carries an odd $Z_{2}^{\prime }$ charge whereas the other dark
matter component carries an odd-$Z_{2}$ charge. As a result, the particle
spectrum in the model can be separated into two parts: the DM sector, which
contains at least one of two odd-$Z_{2},Z_{2}^{\prime }$ charges, and the
active sector, which carries both even-$Z_{2},Z_{2}^{\prime }$ charges.
Although the explicit CP violation is in the dark sector, the preserved $%
Z_{2}\times Z_{2}^{\prime }$ symmetry prohibits the mixing between the
active and dark sectors, then implying that the number of CP violations is
not constrained by the SM Higgs couplings. 
\begin{table}[tbp]
\begin{tabular}{|c|c|c|c|c|c|}
\hline
& $SU\left( 3\right) _{C}$ & $SU\left( 2\right) _{L}$ & $U\left( 1\right)
_{Y}$ & $U\left( 1\right) _{X}$ & $Z_{2}$ \\ \hline
{$\phi $} & $1$ & $2$ & $\frac{1}{2}$ & $0$ & $0$ \\ \hline
$\eta $ & $1$ & $2$ & $\frac{1}{2}$ & $-1$ & $0$ \\ \hline
$\sigma $ & $1$ & $1$ & $0$ & $-2$ & $0$ \\ \hline
$\varphi _{1}$ & $1$ & $1$ & $0$ & $1$ & $1$ \\ \hline
$\varphi _{2}$ & $1$ & $1$ & $0$ & $1$ & $0$ \\ \hline
$\varphi _{3}$ & $1$ & $1$ & $0$ & $-2$ & $1$ \\ \hline
\end{tabular}
\caption{Scalar assignments under $SU\left( 3\right) _{C}\times SU\left(
2\right) _{L}\times U\left( 1\right) _{Y}\times U\left( 1\right) _{X}\times
Z_{2}$.}
\label{scalars}
\end{table}
\begin{table}[tbp]
\begin{tabular}{|c|c|c|c|c|c|}
\hline
& $SU\left( 3\right) _{C}$ & $SU\left( 2\right) _{L}$ & $U\left( 1\right)
_{Y}$ & $U\left( 1\right) _{X}$ & $Z_{2}$ \\ \hline
$q_{iL}$ & $3$ & $2$ & $\frac{1}{6}$ & $0$ & $0$ \\ \hline
$u_{3R}$ & $3$ & $1$ & $\frac{2}{3}$ & $0$ & $0$ \\ \hline
$u_{nR}$ & $3$ & $1$ & $\frac{2}{3}$ & $2$ & $0$ \\ \hline
$d_{3R}$ & $3$ & $1$ & $-\frac{1}{3}$ & $0$ & $0$ \\ \hline
$d_{nR}$ & $3$ & $1$ & $-\frac{1}{3}$ & $-2$ & $0$ \\ \hline
$T_{nL}$ & $3$ & $1$ & $\frac{2}{3}$ & $1$ & $0$ \\ \hline
$T_{nR}$ & $3$ & $1$ & $\frac{2}{3}$ & $-1$ & $0$ \\ \hline
$B_{nL}$ & $3$ & $1$ & $-\frac{1}{3}$ & $-1$ & $0$ \\ \hline
$B_{nR}$ & $3$ & $1$ & $-\frac{1}{3}$ & $1$ & $0$ \\ \hline
\end{tabular}%
\caption{Quark assignments under $SU\left( 3\right) _{C}\times SU\left(
2\right) _{L}\times U\left( 1\right) _{Y}\times U\left( 1\right) _{X}\times
Z_{2}$. Here $i=1,2,3$ and $n=1,2$.}
\label{quarks}
\end{table}

\begin{table}[tbp]
\begin{tabular}{|c|c|c|c|c|c|}
\hline
& $SU\left( 3\right) _{C}$ & $SU\left( 2\right) _{L}$ & $U\left( 1\right)
_{Y}$ & $U\left( 1\right) _{X}$ & $Z_{2}$ \\ \hline
$l_{iL}$ & $1$ & $2$ & $-\frac{1}{2}$ & $-4 $ & $0$ \\ \hline
$l_{nR}$ & $1$ & $1$ & $-1$ & $-2$ & $0$ \\ \hline
$l_{3R}$ & $1$ & $1$ & $-1$ & $-4$ & $0$ \\ \hline
$E_{nL}$ & $1$ & $1$ & $-1$ & $-1$ & $0$ \\ \hline
$E_{nR}$ & $1$ & $1$ & $-1$ & $-3$ & $0$ \\ \hline
$\nu _{nR}$ & $1$ & $1$ & $0$ & $-4$ & $0$ \\ \hline
$N_{nR}$ & $1$ & $1$ & $0$ & $2$ & $0$ \\ \hline
$\Psi _{nR}$ & $1$ & $1$ & $0$ & $-1$ & $1$ \\ \hline
$\Omega _{nR}$ & $1$ & $1$ & $0$ & $-1$ & $0$ \\ \hline
\end{tabular}%
\caption{Lepton assignments under $SU\left( 3\right) _{C}\times SU\left(
2\right) _{L}\times U\left( 1\right) _{Y}\times U\left( 1\right) _{X}\times
Z_{2}$. Here $i=1,2,3$ and $n=1,2$.}
\label{leptons}
\end{table}

With the particle content and symmetries specified in Tab. \ref{scalars}, %
\ref{quarks} and \ref{leptons}, the following quark and leptonic Yukawa
terms arise: 
\begin{eqnarray}
-\mathcal{L}_{Y}^{\left( q\right) } &=&\sum_{i=1}^{3}y_{i}^{\left( u\right) }%
\overline{q}_{iL}\widetilde{\phi }u_{3R}+\sum_{i=1}^{3}%
\sum_{k=1}^{2}x_{ik}^{\left( T\right) }\overline{q}_{iL}\widetilde{\eta }%
T_{kR}+\sum_{k=1}^{2}\sum_{n=1}^{2}z_{kn}^{\left( u\right) }\overline{T}%
_{kL}\varphi _{2}^{\ast }u_{nR} + \sum_{k=1}^{2} \kappa^{(u)}_{k3}\overline{T%
}_{kL} \varphi_2 u_{3R}  \notag \\
&&+\sum_{k=1}^{2}\sum_{s=1}^{2}y_{ks}^{\left( T\right) }\overline{T}%
_{kL}\sigma ^{\ast }T_{sR} +\sum_{i=1}^{3}y_{i}^{\left( d\right) }\overline{q%
}_{iL}\phi d_{3R}+\sum_{i=1}^{2}\sum_{k=1}^{2}x_{ik}^{\left( B\right) }%
\overline{q}_{iL}\eta B_{kR}  \notag \\
&&+\sum_{k=1}^{2}\sum_{n=1}^{2}z_{kn}^{\left( d\right) }\overline{B}%
_{kL}\varphi _{2}d_{nR} + \sum_{k=1}^{2} \kappa^{(d)}_{k3}\overline{B}_{kL}
\varphi_2^* d_{3R} +\sum_{k=1}^{2}\sum_{n=1}^{2}y_{kn}^{\left( B\right) }%
\overline{B}_{kL}\sigma B_{nR}  \label{Yukawa1a}
\end{eqnarray}

\begin{eqnarray}
-\mathcal{L}_{Y}^{\left( l\right) } &=&\sum_{i=1}^{3}y_{i3}^{\left( l\right)
}\overline{l}_{iL}\phi l_{3R}+\sum_{i=1}^{3}\sum_{k=1}^{2}x_{ik}^{\left(
E\right) }\overline{l}_{iL}\eta
E_{kR}+\sum_{n=1}^{2}\sum_{k=1}^{2}z_{nk}^{\left( l\right) }\overline{E}%
_{nL}\varphi _{2}l_{kR}+\sum_{n=1}^{2}\sum_{k=1}^{2}y_{nk}^{\left( E\right) }%
\overline{E}_{nL}\sigma ^{\ast }E_{kR}  \notag \\
&&+\sum_{i=1}^{3}\sum_{n=1}^{2}\left( y_{\nu }\right) _{in}\overline{l}_{iL}%
\widetilde{\phi }\nu _{nR}+\sum_{n=1}^{2}\sum_{k=1}^{2}\left( y_{N}\right)
_{nk}\nu _{nR}\sigma ^{\ast }\overline{N_{kR}^{C}}+\sum_{n=1}^{2}%
\sum_{k=1}^{2}\left( y_{\Psi }\right) _{nk}\overline{N}_{nR}\Psi
_{kR}^{C}\varphi _{1}  \notag \\
&&+\sum_{n=1}^{2}\sum_{k=1}^{2}\left( x_{\Omega }\right) _{nk}\overline{\Psi 
}_{nR}\varphi _{3}\Omega _{kR}^{C}+\sum_{n=1}^{2}\sum_{k=1}^{2}\left(
y_{\Omega }\right) _{nk}\overline{\Omega }_{nR}\sigma \Omega _{kR}^{C}+h.c.
\label{Yuka2}
\end{eqnarray}

\subsection{Scalar potential}

With the scalar content and symmetries specified in Tab.(\ref%
{scalars}), the following scalar potential arises: 
\begin{eqnarray}
\mathcal{V}= V+V_{\text{CPV}} +V^{\prime}_{\text{CPV}}
\end{eqnarray}
where 
\begin{eqnarray}
V=&& -\mu_\phi^2 \phi^\dag \phi -\mu_\eta^2 \eta^\dag \eta -\mu_\sigma^2
\sigma^\dag \sigma-\sum_{i=1}^3\mu^2_{\varphi_i} \varphi_i^\dag \varphi_i +%
\frac{\lambda_\phi}{4}( \phi^\dag \phi)^2+\frac{\lambda_\eta}{4} ( \eta^\dag
\eta)^2+\frac{\lambda_\sigma}{4}(\sigma^\dag \sigma)^2+\sum_{i=1}^3 \frac{%
\lambda_{\varphi_i}}{4}(\varphi_i^\dag \varphi_i)^2  \notag \\
&&+\lambda_{\eta \phi}(\phi^\dag \phi)(\eta^\dag \eta)+\lambda^\prime_{\eta
\phi} (\eta^\dag \phi)(\phi^\dag \eta) +\lambda_{\sigma \phi}(\sigma^\dag
\sigma)(\phi^\dag \phi)+\lambda_{\sigma \eta}(\sigma^\dag \sigma)(\eta^\dag
\eta)+\sum_{i=1}^3\lambda_{\sigma \varphi_i}(\sigma^\dag
\sigma)(\varphi_i^\dag \varphi_i)  \notag \\
&& +\sum_{i=1}^3 \left\{\lambda_{\phi \varphi_i}(\phi^\dag
\phi)(\varphi_i^\dag \varphi_i) +\lambda_{\eta \varphi_i}(\eta^\dag
\eta)(\varphi_i^\dag \varphi_i)\right\} +\sum_{i\neq j;
i,j=1}^3\lambda_{\varphi_i \varphi_j} (\varphi_i
\varphi_j^\dag)(\varphi_i^\dag \varphi_j) ,
\end{eqnarray}
and 
\begin{eqnarray}
V_{\text{CPV}} && = \left\{\mathcal{F}_\varphi \varphi_1 \varphi_2 \varphi_3+%
\mathcal{G} \eta^\dag \phi \varphi_2^\dag +\mathcal{F}_{\sigma \varphi_1}
\sigma \varphi_1 \varphi_1+\mathcal{F}_{\sigma \varphi_2} \sigma \varphi_2
\varphi_2+H.c. \right\},  \notag \\
V_{\text{CPV}}^\prime && = \left\{f\eta^\dag \phi
\varphi_1\varphi_3+f^\prime\eta^\dag \phi \sigma \varphi_2+f_\sigma
\sigma^\dag \varphi_3 \varphi^\dag_2 \varphi_1+f_\sigma^\prime \sigma^\dag
\varphi_3 \varphi^\dag_1 \varphi_2+f_{\varphi_3\sigma} \varphi_3^2
\sigma^{*2}+f_{\varphi_1 \varphi_2}\varphi_1^2 \varphi_2^{*2} +H.c. \right\},
\label{Higgspotential2}
\end{eqnarray}
All parameters in the potential $V$ are real, whereas those in the
CP-violating parts, $V_{\text{CPV}}$ and $V_{\text{CPV}}^\prime$, can be
complex. However, by exploiting the freedom of rephasing the scalar fields,
the complex phases can be absorbed into $\eta, \varphi_1, \varphi_2,
\varphi_3, \sigma,$ and $\phi$, making $\mathcal{F}{\sigma \varphi_1}, 
\mathcal{F}{\sigma \varphi_2}, f, f^\prime, f_\sigma,$ and $%
f_{\sigma^\prime} $ real. The remaining complex phases in $\mathcal{F}%
_{\varphi}, \mathcal{G}, f_{\varphi_3 \sigma},$ and $f_{\varphi_1 \varphi_2}$
introduce CP violation in the scalar potential, thereby making the model
explicitly CP-violating and generating complex corrections to the quark
masses at the radiative level.\newline

\subsubsection{Theoretical constraints on the scalar potential}

To ensure the validity of the effective description, we require the scalar
sector to satisfy three sets of theoretical bounds: perturbativity, vacuum
stability, and tree--level unitarity.

\paragraph{Perturbativity.}

All dimensionless couplings in the scalar potential are constrained to the perturbative regime.
\begin{equation}
|\lambda_i|,\, |f_j| \;<\; 4\pi ,
\end{equation}
where $\lambda_i$ denote the quartic couplings and $f_j$ represent the
dimensionless coefficients of the CP--violating interactions. This condition
guarantees the convergence of perturbative expansions.

\paragraph{Vacuum stability.}

The quartic part of the scalar potential, which dominate the behavior of the scalar potential in the region of
very large values of the field components, must be bounded from below in all
field directions. Following the procedure used for analyzing the stability described in Refs \cite{Maniatis:2006fs,Bhattacharyya:2015nca}, we find
that all self--couplings are positive,
\begin{equation}
\lambda_\phi,\,\lambda_\eta,\,\lambda_\sigma,\,\lambda_{\varphi_i} \;>\; 0,
\end{equation}
together with copositivity constraints on the portal couplings, 
\begin{align}
\lambda_{\phi\eta} + \sqrt{\lambda_\phi \lambda_\eta} &> 0, & 
\lambda_{\phi\sigma} + \sqrt{\lambda_\phi \lambda_\sigma} &> 0, & 
\lambda_{\eta\sigma} + \sqrt{\lambda_\eta \lambda_\sigma} &> 0, \\
\lambda_{\phi\varphi_i} + \sqrt{\lambda_\phi \lambda_{\varphi_i}} &> 0, & 
\lambda_{\eta\varphi_i} + \sqrt{\lambda_\eta \lambda_{\varphi_i}} &> 0, & 
\lambda_{\sigma\varphi_i} + \sqrt{\lambda_\sigma \lambda_{\varphi_i}} &> 0,
\qquad (i=1,2,3).
\end{align}
and analogous inequalities for the mixed couplings $\lambda_{\varphi_i%
\varphi_j}$. These constraints ensure that the potential does not develop
runaway directions with $V\to -\infty$.

It is worth noting that the operator $\lambda_{\eta\phi}^\prime(\eta^\dag%
\phi)(\phi^\dag\eta)$ does not in general reduce to $(\eta^\dag\eta)(\phi^%
\dag\phi)$. To account for this phase dependence, one may define an
effective portal coupling 
\begin{equation}
\lambda_{\eta\phi}^{\text{eff}} = \lambda_{\eta\phi} +
\min\!\left(0,\lambda^\prime_{\eta\phi}\right),
\end{equation}
and use $\lambda_{\eta\phi}\to\lambda_{\eta\phi}^{\text{eff}}$ in the
pairwise conditions above. This provides a conservative, sufficient
criterion for vacuum stability.

\paragraph{Tree-level unitarity.}

The unitarity of $2\to2$ scalar scattering amplitudes imposes additional
constraints on the quartic scalar couplings. In particular, the $s$--wave component
of the amplitude must satisfy 
\begin{equation}
|\mathrm{Re}\, a_0| < \tfrac{1}{2},
\end{equation}
which translates into bounds on linear combinations of the couplings.
Operationally, one constructs the scattering matrix $\mathcal{M}$ for all
two--body scalar states, diagonalizes it, and imposes the condition 
\begin{equation}
|\Lambda_i| < 8\pi ,
\end{equation}
where $\Lambda_i$ are the eigenvalues of $\mathcal{M}$.


\medskip 
When combined with vacuum stability and perturbativity, the unitarity constraints significantly restrict the parameter space, ensuring the scalar potential remains consistent and theoretically controlled across the relevant energy scales.

\subsubsection{Impact of CP-violating terms on theoretical bounds}

The potential includes CP--violating contributions $V_{\text{CPV}}$ and $V_{%
\text{CPV}}^{\prime }$ with cubic and quartic interactions. The theoretical constraints on the CP violating scalar couplings are described below: 

\paragraph{Perturbativity.}

All CP--violating couplings $f_j$ must remain perturbative, 
\begin{equation}
|f_j| < 4\pi .
\end{equation}

\paragraph{Vacuum stability.}

Cubic terms are irrelevant at large field values, while quartic
CP--violating couplings contribute to the vacuum stability conditions. The
bounded from below conditions 
apply to their real parts, as follows: 
\begin{equation}
\mathrm{Re}\,f_{\varphi_3\sigma} > -\sqrt{\lambda_{\varphi_3}\lambda_\sigma}%
, \qquad \mathrm{Re}\,f_{\varphi_1\varphi_2} > -\sqrt{\lambda_{\varphi_1}%
\lambda_{\varphi_2}} .
\end{equation}



\medskip In short, CP--violating terms extend the bounds: their real parts
enter in the vacuum stability conditions, and their magnitudes are constrained by
perturbativity.

\subsubsection{Scalar mass spectrum}

The global minimum conditions for the scalar potential yield the following
solutions for the mass dimensionful parameters $\mu^2_{\phi}$ and $%
\mu^2_{\sigma}$: 
\begin{eqnarray}
\mu^2_{\phi} =\frac{1}{2} \lambda _{\sigma \phi } v_{\sigma }^2 +\frac{1}{4}
\lambda _{\phi } v_{\phi }^2, \hspace{1cm} \mu^2_{\sigma} = \frac{1}{2}
\lambda _{\sigma \phi } v_{\phi }^2+\frac{1}{4} \lambda _{\sigma } v_{\sigma
}^2.
\end{eqnarray}
The charged scalar fields are found in the first components of the scalar
doublets, denoted as $\phi$ and $\eta$. The first component of $\phi$,
referred to as $\phi^\pm$, is massless and is identified as the Goldstone
bosons that are "eaten" by the longitudinal components of the $W^\pm$ gauge
bosons. In contrast, the first component of $\eta$, called $\eta^\pm$,
corresponds to physical fields that have mass 
\begin{eqnarray}
m^2_{\eta^\pm}=-\mu _{\eta }^2+\frac{1}{2} \lambda _{\eta \phi } v_{\phi }^2+%
\frac{1}{2} \lambda _{\sigma \eta } v_{\sigma }^2 .
\end{eqnarray}
Two CP-even neutral scalar fields, both carrying even $Z_2$ and $Z_2^{\prime
}$ charges, $\Re \phi$ and $\Re \sigma$, mix via a $2 \times 2$ matrix. We
find the physical states for them as follows: 
\begin{eqnarray}
h= \cos \theta_h \Re \phi-\sin \theta_h \Re \sigma, \hspace{0.5cm} H=\sin
\theta_h \Re \phi+ \cos \theta_h \Re \sigma
\end{eqnarray}
with $\tan 2 \theta_h= \frac{4\lambda_{\sigma \phi}v_\sigma v_\phi }{%
\lambda_\sigma v_\sigma^2-\lambda_\phi v_\phi^2}$, and their masses are
respectively given by: 
\begin{eqnarray}
m_h^2 &=& \frac{1}{4}\left( \lambda_\phi v_\phi^2+\lambda_\sigma v_\sigma^2-%
\sqrt{ \lambda_\phi ^2v_\phi^4+\lambda_\sigma^2 v_\sigma^4-2(\lambda_\phi
\lambda_\sigma-8\lambda_{\sigma \phi})v_\phi^2v_\sigma^2}\right) ,  \notag \\
m_H^2&=&\frac{1}{4}\left( \lambda_\phi v_\phi^2+\lambda_\sigma v_\sigma^2+%
\sqrt{ \lambda_\phi ^2v_\phi^4+\lambda_\sigma^2 v_\sigma^4-2(\lambda_\phi
\lambda_\sigma-8\lambda_{\sigma \phi})v_\phi^2v_\sigma^2}\right) ,
\end{eqnarray}
where $h$ is identified as the $126$ GeV SM-like Higgs boson, and $H$ is a
new CP-even Higgs boson. In contrast, two CP-odd neutral scalar fields, both
carrying even $Z_2$ and $Z_2^{\prime }$ charges, $\Im \phi$ and $\Im \sigma$%
, are massless, and interpreted as the Goldstone boson eaten by the
longitudinal component of the $Z$ gauge boson and the Majoron, respectively.
In the dark scalar sector of the model there is the field, $\varphi_1$,
which carries both odd $Z_2$ and $Z_2^{\prime }$ charges. Due to the
interaction terms between $\varphi_1$, $\phi$, and $\sigma$, $\Re \varphi_1$
and $\Im \varphi_1$ are physical dark CP even and CP odd scalars,
respectively, with masses given by: 
\begin{eqnarray}
m^2_{\Re \varphi_1}&=& -\mu_{\varphi_1}^2+\frac{1}{2}\lambda_{\phi
\varphi_1}v_\phi^2+\frac{1}{2}\lambda_{\sigma \varphi_1}v^2_\sigma +\sqrt{2}%
\mathcal{F}_{\sigma \phi_1}v_\sigma,  \notag \\
m^2_{\Im \varphi_1}&=& -\mu_{\varphi_1}^2+\frac{1}{2}\lambda_{\phi
\varphi_1}v_\phi^2+\frac{1}{2}\lambda_{\sigma \varphi_1}v^2_\sigma -\sqrt{2}%
\mathcal{F}_{\sigma \phi_1}v_\sigma.
\end{eqnarray}
On the other hand, the dark singlet scalar field $\varphi_3$ has an odd $Z_2$
charge and is neutral under the preserved $Z_2^{\prime }$ symmetry. 
After symmetry breaking, the model predicts a mixing matrix between the real
and imaginary components of $\varphi_3 $, which arises from the explicit
CP-breaking term $f_{\varphi_3 \sigma}\varphi_3^2 \sigma^{*2}+H.c.$. The physical
states are mixtures of the CP-odd and CP-even components, defined as
follows: 
\begin{eqnarray}
R_{\varphi_3}= \cos \theta_{\varphi_3} \Re \varphi_3 - \sin
\theta_{\varphi_3} \Im \varphi_3, \hspace{0.5cm} I_{\varphi_3}=\sin
\theta_{\varphi_3} \Re \varphi_3+ \cos \theta_{\varphi_3} \Im \varphi_3
\end{eqnarray}
where the mixing angle satisfies the relation $\tan \theta_{\varphi_3}= 
\frac{\Im f_{\varphi_3 \sigma} }{\Re f_{\varphi_3 \sigma }}$. The masses of these states
are given by: 
\begin{eqnarray}
m^2_{R_{\varphi_3}}&=&-\mu_{\varphi_3}^2+\frac{\lambda_{\phi \varphi_3}}{2}%
v_\phi^2+\frac{\lambda_{\sigma \varphi_3}}{2}v_{\sigma}^2+|f_{\varphi_3 \sigma}|v_{%
\sigma}^2,  \notag \\
m^2_{I_{\varphi_3}}&=&-\mu_{\varphi_3}^2+\frac{\lambda_{\phi \varphi_3}}{2}%
v_\phi^2+\frac{\lambda_{\sigma \varphi_3}}{2}v_{\sigma}^2-|f_{\varphi_3 \sigma}|v_{%
\sigma}^2.
\end{eqnarray}
The neutral scalar fields that carry an even $Z_2$ charge and an odd $%
Z_2^\prime$ correspond to the second component of the scalar doublet $\eta$,
denoted as $\eta^0$, and to $\varphi_2$. Due to the complex phases
associated with $\mathcal{G}$ and $f^\prime$, there is a mixing between the
CP-even components $(\Re \eta, \Re \varphi_2)$ and the CP-odd components $%
(\Im \eta, \Im \varphi_2)$. The mixing matrix has a following form 
\begin{eqnarray}
M^2_{\Re \eta \varphi_2}=\left( 
\begin{array}{cc}
M^2_{\Re \eta \varphi_2} & M^2_{\Re \Im \eta \varphi_2} \\ 
M^2_{\Re \Im \eta \varphi_2} & M^2_{ \Im \eta \varphi_2} \\ 
& 
\end{array}
\right)  \label{mixcp}
\end{eqnarray}
where 
\begin{eqnarray}
M^2_{\Re \eta \varphi_2} &=&\left( 
\begin{array}{cc}
-\mu^2_{\eta}+\frac{1}{2}\left(\lambda_{\eta \phi}
v_{\phi}^2+\lambda^\prime_{\eta \phi} v_{\phi}^2+\lambda_{\sigma
\eta}v_\sigma^2\right) & \frac{1}{2}\left(\sqrt{2}\Re \mathcal{G}+ \Re
f^\prime v_\sigma\right)v_\phi \\ 
\frac{1}{2}\left(\sqrt{2}\Re \mathcal{G}+ \Re f^\prime v_\sigma\right)v_\phi
& -\mu_{\varphi_2}^2+\frac{1}{2}\left( \lambda_{\phi
\varphi_2}v_\phi^2+\lambda_{\sigma \varphi_2}v_\sigma^2+2\sqrt{2}\mathcal{F}%
_{\sigma \varphi_2}v_\sigma\right) \\ 
& 
\end{array}
\right),  \notag \\
M^2_{\Re \Im \eta \varphi_2}&=& \left( 
\begin{array}{cc}
0 & - \frac{v_\phi}{2}\left(\sqrt{2}\Im \mathcal{G}-\Im f^\prime v_\sigma
\right) \\ 
- \frac{v_\phi}{2}\left(\sqrt{2}\Im \mathcal{G}-\Im f^\prime v_\sigma \right)
& 0 \\ 
& 
\end{array}
\right),  \notag \\
M^2_{ \Im \eta \varphi_2} &=&\left( 
\begin{array}{cc}
-\mu_{\eta}^2 +\frac{1}{2}\left(\lambda_{\eta \phi}v_\phi^2 +\lambda_{\eta
\phi}^\prime v_\phi^2+\lambda_{\sigma \eta }v_\sigma^2\right) & \frac{1}{2}%
\left(\sqrt{2}\Re \mathcal{G}- \Re f^\prime v_\sigma\right)v_\phi \\ 
\frac{1}{2}\left(\sqrt{2}\Re \mathcal{G}- \Re f^\prime v_\sigma\right)v_\phi
& -\mu_{\varphi_2}^2+\frac{1}{2}\left( \lambda_{\phi
\varphi_2}v_\phi^2+\lambda_{\sigma \varphi_2}v_\sigma^2-2\sqrt{2}\mathcal{F}%
_{\sigma \varphi_2}v_\sigma \right) \\ 
& 
\end{array}
\right).
\end{eqnarray}
The mixing matrix presented in Eq.(\ref{mixcp}) yields four physical states,
denoted as $\rho^T=\left (\rho_1, \rho_2, \rho_3, \rho_4 \right)$. These
states are related to the states contained in $RI_{\eta \varphi_2}^T=\left(
\Re \eta^0, \Re \varphi_2, \Im \eta^0, \Im \varphi_2 \right)$ through a $4
\times 4$ rotation matrix, referred as $R$. The relationship can be
expressed as $\rho^T= R^{T}RI_{\eta \varphi_2}^T$.

\section{Fermion masses and mixings}

\subsection{Quark masses and mixings}

\label{fermionmassesandmixings} In the model under consideration, the third
quark family acquires its mass at the tree-level, while the other SM quarks
get their masses radiatively at one-loop level. To understand this in
more detail, we will begin by analyzing the quark Yukawa interactions. 
\newline

The up- and down-type quark mass terms can be written in the interaction
basis as follows: 
\begin{equation}
\mathcal{L}^{q}_{\mathrm{mass}} \;=\; 
\begin{pmatrix}
\bar{u}_{1L} & \bar{u}_{2L} & \bar{u}_{3L}%
\end{pmatrix}
M_u 
\begin{pmatrix}
u_{1R} \\ 
u_{2R} \\ 
u_{3R}%
\end{pmatrix}
+ 
\begin{pmatrix}
\bar{d}_{1L} & \bar{d}_{2L} & \bar{d}_{3L}%
\end{pmatrix}
M_d 
\begin{pmatrix}
d_{1R} \\ 
d_{2R} \\ 
d_{3R}%
\end{pmatrix}
+ \mathrm{h.c.},
\end{equation}
where $M_u$ and $M_d$ are $3\times3$ mass matrices. These mass matrices can
be expressed as a sum of contributions from different loop orders: 
\begin{equation*}
M_u = M_u^{(0)} + M_u^{(1)} + M_u^{(2)} + \cdots, \qquad M_d = M_d^{(0)} +
M_d^{(1)} + M_d^{(2)} + \cdots .
\end{equation*}
At the tree level, only the terms in the third column are nonzero, which imposes a specific structure on the matrices:
\begin{equation}
M_u^{(0)} = 
\begin{pmatrix}
0 & 0 & \tfrac{y_1^{(u)}}{\sqrt{2}}v \\ 
0 & 0 & \tfrac{y_2^{(u)}}{\sqrt{2}}v \\ 
0 & 0 & \tfrac{y_3^{(u)}}{\sqrt{2}}v%
\end{pmatrix}%
, \qquad M_d^{(0)} = 
\begin{pmatrix}
0 & 0 & \tfrac{y_1^{(d)}}{\sqrt{2}}v \\ 
0 & 0 & \tfrac{y_2^{(d)}}{\sqrt{2}}v \\ 
0 & 0 & \tfrac{y_3^{(d)}}{\sqrt{2}}v%
\end{pmatrix}%
.
\end{equation}

\begin{figure}[]
\vspace{0cm} \centering
\includegraphics[width=0.8\textwidth]{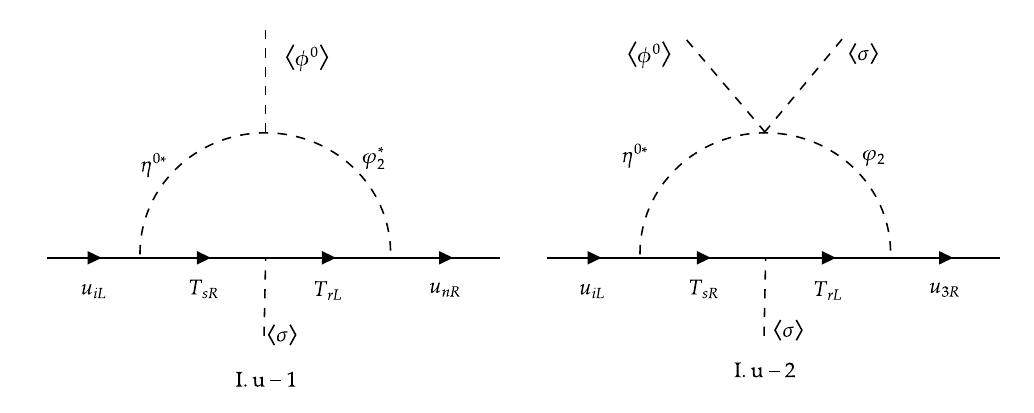}// \includegraphics[width=0.8
\textwidth]{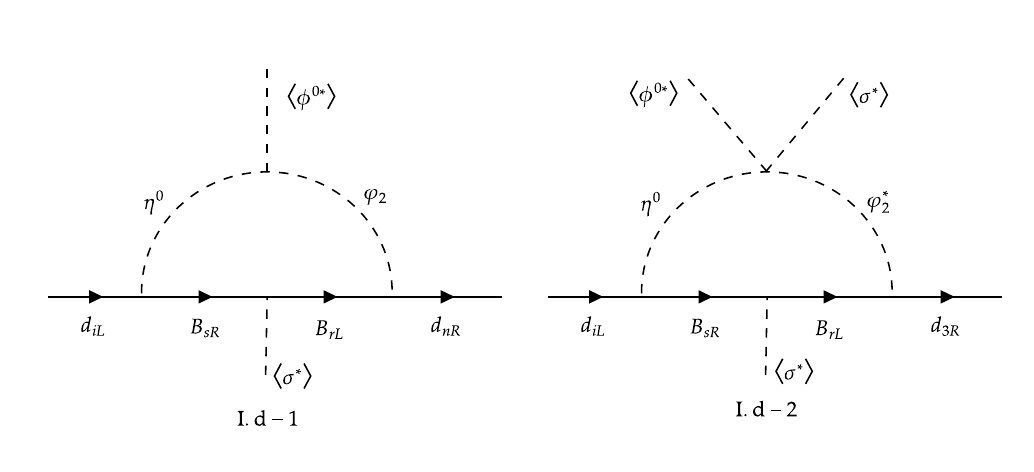} \vspace{0cm}
\caption{ One-loop corrections to the SM u-quark}
\label{Quark-oneloop}
\end{figure}

When one-loop corrections are considered, they introduce additional nonzero
elements. The one-loop diagrams contributing to the quark mass matrix are
illustrated in Fig. \ref{Quark-oneloop}. 

We enforce an exact CP symmetry within the SM part of the Lagrangian, but
allow for explicit CP breaking in the dark sector. This ensures that CP
phases vanish at tree level. We observe that the CP-violating phase $%
(e^{\pm i \alpha})$ in the dark matter sector, arising from the complex coupling $\mathcal{G}$ of the scalar interaction vertex 
$\mathcal{G} \eta^\dag \phi \varphi_2^\dag +h.c$, induces
corrections to the first and second columns of the mass mixing matrix, while
leaving the third column unaffected. The full mass matrices, up to one-loop
order, are given by the sum of the tree-level and one-loop contributions, $%
M^{(0)}+M^{(1)}$. This results in the following structure: 
\begin{equation}
\mathcal{M}_u = 
\begin{pmatrix}
e^{i\alpha} \left( m_{11}^{u} \right)^{(1)} & e^{i\alpha} \left( m_{12}^{u}
\right)^{(1)} & \tfrac{y_1^{(u)}}{\sqrt{2}}v + \left( m_{13}^{u}
\right)^{(1)} \\[4pt] 
e^{i\alpha}\left( m_{21}^{u} \right)^{(1)} & e^{i\alpha}\left( m_{22}^{u}
\right)^{(1)} & \tfrac{y_2^{(u)}}{\sqrt{2}}v + \left( m_{23}^{u}
\right)^{(1)} \\[4pt] 
e^{i\alpha} \left( m_{31}^{u} \right)^{(1)} & e^{i\alpha} \left( m_{32}^{u}
\right)^{(1)} & \tfrac{y_3^{(u)}}{\sqrt{2}}v + \left( m_{33}^{u}
\right)^{(1)}%
\end{pmatrix}%
, \qquad \mathcal{M}_d = 
\begin{pmatrix}
e^{-i\alpha} \left( m_{11}^{d} \right)^{(1)} & e^{-i\alpha} \left(
m_{12}^{d} \right)^{(1)} & \tfrac{y_1^{(d)}}{\sqrt{2}}v + \left( m_{13}^{d}
\right)^{(1)} \\[4pt] 
e^{-i\alpha} \left( m_{21}^{d} \right)^{(1)} & e^{-i\alpha} \left(
m_{22}^{d} \right)^{(1)} & \tfrac{y_2^{(d)}}{\sqrt{2}}v + \left( m_{23}^{d}
\right)^{(1)} \\[4pt] 
e^{-i\alpha} \left( m_{31}^{d} \right)^{(1)} & e^{-i\alpha} \left(
m_{32}^{d} \right)^{(1)} & \tfrac{y_3^{(d)}}{\sqrt{2}}v + \left( m_{33}^{d}
\right)^{(1)}%
\end{pmatrix}%
.  \label{eq:udquarkmatrices}
\end{equation}
Here, the entries 
$\left( m_{ij}^{q} \right)^{(1)}$ ($i,j=1,2,3$) with $q=u,d$, which arise
from one-loop corrections, have the following form 
\begin{eqnarray}
e^{i\alpha } \left( m_{in}^{u} \right)^{(1)} &=&\dsum\limits_{r=1}^{2}\frac{%
x_{ir}^{\left( T\right) }z_{rn}^{\left( u\right) }m_{T_{r}}}{16\pi ^{2}}%
\left\{ \sum_{i=1}^{4}R_{1i}R_{2i}f\left( m_{\rho
_{i}}^{2},m_{T_{r}}^{2}\right) -\sum_{i=1}^{4}R_{3i}R_{4i}f\left( m_{\rho
_{i}}^{2},m_{T_{r}}^{2}\right) \right\}  \notag \\
&&-i\dsum\limits_{r=1}^{2}\frac{x_{ir}^{\left( T\right) }z_{rn}^{\left(
u\right) }m_{T_{r}}}{16\pi ^{2}}\left\{ \sum_{i=1}^{4}R_{1i}R_{4i}f\left(
m_{\rho _{i}}^{2},m_{T_{r}}^{2}\right) +\sum_{i=1}^{4}R_{2i}R_{3i}f\left(
m_{\rho _{i}}^{2},m_{T_{r}}^{2}\right) \right\}  \label{eq:mupquarkentries}
\end{eqnarray}%
\begin{eqnarray}
e^{-i\alpha} \left( m_{in}^{d} \right)^{(1)} &=&\dsum\limits_{r=1}^{2}\frac{%
x_{ir}^{\left( B\right) }z_{rn}^{\left( d\right) }m_{B_{r}}}{16\pi ^{2}}%
\left\{ \sum_{i=1}^{4}R_{1i}R_{2i}f\left( m_{\rho
_{i}}^{2},m_{B_{r}}^{2}\right) -\sum_{i=1}^{4}R_{3i}R_{4i}f\left( m_{\rho
_{i}}^{2},m_{B_{r}}^{2}\right) \right\}  \notag \\
&&+i\dsum\limits_{r=1}^{2}\frac{x_{ir}^{\left( B\right) }z_{rn}^{\left(
d\right) }m_{B_{r}}}{16\pi ^{2}}\left\{ \sum_{i=1}^{4}R_{1i}R_{4i}f\left(
m_{\rho _{i}}^{2},m_{B_{r}}^{2}\right) +\sum_{i=1}^{4}R_{2i}R_{3i}f\left(
m_{\rho _{i}}^{2},m_{B_{r}}^{2}\right) \right\}  \label{quark2}
\end{eqnarray}
and 
\begin{eqnarray}
\left( m_{i3}^{u} \right)^{(1)} &=&\dsum\limits_{r=1}^{2}\frac{%
x_{ir}^{\left( T\right) }\kappa_{r3}^{\left( u\right) }m_{T_{r}}}{16\pi ^{2}}%
\left\{ \sum_{i=1}^{4}R_{1i}R_{2i}f\left( m_{\rho
_{i}}^{2},m_{T_{r}}^{2}\right) -\sum_{i=1}^{4}R_{3i}R_{4i}f\left( m_{\rho
_{i}}^{2},m_{T_{r}}^{2}\right) \right\}  \notag \\
&&-i\dsum\limits_{r=1}^{2}\frac{x_{ir}^{\left( T\right) }\kappa_{r3}^{\left(
u\right) }m_{T_{r}}}{16\pi ^{2}}\left\{ \sum_{i=1}^{4}R_{1i}R_{4i}f\left(
m_{\rho _{i}}^{2},m_{T_{r}}^{2}\right) +\sum_{i=1}^{4}R_{2i}R_{3i}f\left(
m_{\rho _{i}}^{2},m_{T_{r}}^{2}\right) \right\}  \label{eq:mupquarkentries}
\end{eqnarray}%
\begin{eqnarray}
\left(m_{i3}^{d} \right)^{(1)} &=&\dsum\limits_{r=1}^{2}\frac{x_{ir}^{\left(
B\right) }\kappa_{r3}^{\left( d\right) }m_{B_{r}}}{16\pi ^{2}}\left\{
\sum_{i=1}^{4}R_{1i}R_{2i}f\left( m_{\rho _{i}}^{2},m_{B_{r}}^{2}\right)
-\sum_{i=1}^{4}R_{3i}R_{4i}f\left( m_{\rho _{i}}^{2},m_{B_{r}}^{2}\right)
\right\}  \notag \\
&&+i\dsum\limits_{r=1}^{2}\frac{x_{ir}^{\left( B\right) }\kappa_{r3}^{\left(
d\right) }m_{B_{r}}}{16\pi ^{2}}\left\{ \sum_{i=1}^{4}R_{1i}R_{4i}f\left(
m_{\rho _{i}}^{2},m_{B_{r}}^{2}\right) +\sum_{i=1}^{4}R_{2i}R_{3i}f\left(
m_{\rho _{i}}^{2},m_{B_{r}}^{2}\right) \right\}  \label{quark2}
\end{eqnarray}
with $i=1,2,3$ and $r,n=1,2$ and the loop function $f\left(
m_{1},m_{2}\right) $ is given by: 
\begin{equation}
f\left( m_{1}^{2},m_{2}^{2}\right) =\frac{m_{1}^{2}}{m_{1}^{2}-m_{2}^{2}}\ln
\left( \frac{m_{1}^{2}}{m_{2}^{2}}\right).  \label{eq:1loopfunction}
\end{equation}
The phase factor $e^{\pm i\alpha}$ acts as a crucial link between the dark
and visible sectors. It encapsulates the CP-breaking effects originating in
the dark sector and transmits them to the quark mass matrices at the
one-loop level. Through this mechanism, the model naturally accounts for the
observed CP violation in the weak interactions, while the strong CP phase
remains parametrically suppressed. To determine the quantitative value of
the weak phase, we shall conduct a systematic analysis.

For simplicity, we consider a benchmark scenario where $m_{T_{r}}=m_{B_{r}}$ (%
$r=1,2$), which implies $\alpha_{u}=\alpha_{d}=\alpha$. Since the explicit
breaking of the CP symmetry in the dark sector is triggered only by specific
scalar interaction term $\mathcal{G} \eta^\dag \phi \varphi_2^\dag +h.c$, the
parameters $m_{kn}^{u(d)}$ ($k=1,2,3$, $n=1,2$) are taken to be real
numbers. The mass matrices, $M_{u(d)}$, are diagonalized by the usual
bi-unitary transformations $V_{L,R}^{u(d)}$, given by:
\begin{eqnarray}
V_{L}^{d\dag }M_d V_{R}^{d} &=&Diag(m_{d},m_{s},m_{b}),  \notag \\
V_{L}^{u\dag }M_uV_{R}^{u} &=&Diag(m_{u},m_{c},m_{t}),
\end{eqnarray}%
and the CKM quark mixing matrix is determined as $V_{\text{CKM}}=\left(V_{L}^{u}\right)^%
\dag V_{L}^{d}$. The bi-unitary matrices can be parameterized as follows 
\begin{eqnarray}
V_{L}^{u} &=&\mathcal{O}_{L}^{u}Diag\left( 
\begin{array}{ccc}
e^{i\beta _{1}^{u}} & e^{i\beta _{2}^{u}} & e^{i\beta _{3}^{u}}%
\end{array}%
\right),\hspace{1cm}\space\space\space V_{R}^{u}=\mathcal{O}%
_{R}^{u}Diag\left( 
\begin{array}{ccc}
e^{i\left( \beta _{1}^{u}-\alpha \right) } & e^{i\left( \beta
_{2}^{u}-\alpha \right) } & e^{i\beta _{3}^{u}}%
\end{array}%
\right) ,  \notag \\
V_{L}^{d} &=&\mathcal{O}_{L}^{d}Diag\left( 
\begin{array}{ccc}
e^{-i\beta _{1}^{d}} & e^{-i\beta _{2}^{d}} & e^{-i\beta _{3}^{d}}%
\end{array}%
\right),\hspace{1cm}\space\space\space V_{R}^{d}=\mathcal{O}%
_{R}^{d}Diag\left( 
\begin{array}{ccc}
e^{-i\left( \beta _{1}^{d}-\alpha \right) } & e^{-i\left( \beta
_{2}^{d}-\alpha \right) } & e^{-i\left( \beta _{3}^{d}-\alpha \right) }%
\end{array}%
\right),
\end{eqnarray}%
where $\mathcal{O}_{L,R}^{u\left( d\right) }$ are real orthogonal matrices
which are approximatelly given by: 
\begin{eqnarray}
\mathcal{O}_{L}^{u} &=&\left( 
\begin{array}{ccc}
\left( \mathcal{O}_{L}^{u}\right) _{11} & 0 & \left( \mathcal{O}%
_{L}^{u}\right) _{13} \\ 
0 & \left( \mathcal{O}_{L}^{u}\right) _{22} & \left( \mathcal{O}%
_{L}^{u}\right) _{23} \\ 
\left( \mathcal{O}_{L}^{u}\right) _{13} & \left( \mathcal{O}_{L}^{u}\right)
_{23} & \left( \mathcal{O}_{L}^{u}\right) _{33}%
\end{array}%
\right),\hspace{1cm}\space\space\space\mathcal{O}_{R}^{u}=\left( 
\begin{array}{ccc}
\left( \mathcal{O}_{R}^{u}\right) _{11} & \left( \mathcal{O}_{R}^{u}\right)
_{12} & 0 \\ 
\left( \mathcal{O}_{R}^{u}\right) _{12} & \left( \mathcal{O}_{R}^{u}\right)
_{22} & 0 \\ 
0 & 0 & \left( \mathcal{O}_{R}^{u}\right) _{33}%
\end{array}%
\right), \\
\mathcal{O}_{L}^{d} &=&\left( 
\begin{array}{ccc}
\left( \mathcal{O}_{L}^{d}\right) _{11} & 0 & \left( \mathcal{O}%
_{L}^{d}\right) _{13} \\ 
0 & \left( \mathcal{O}_{L}^{d}\right) _{22} & \left( \mathcal{O}%
_{L}^{d}\right) _{23} \\ 
\left( \mathcal{O}_{L}^{d}\right) _{13} & \left( \mathcal{O}_{L}^{d}\right)
_{23} & \left( \mathcal{O}_{L}^{d}\right) _{33}%
\end{array}%
\right),\hspace{1cm}\space\space\space\mathcal{O}_{R}^{d}=\left( 
\begin{array}{ccc}
\left( \mathcal{O}_{R}^{d}\right) _{11} & \left( \mathcal{O}_{R}^{d}\right)
_{12} & 0 \\ 
\left( \mathcal{O}_{R}^{d}\right) _{12} & \left( \mathcal{O}_{R}^{d}\right)
_{22} & 0 \\ 
0 & 0 & \left( \mathcal{O}_{R}^{d}\right) _{33}%
\end{array}%
\right).
\end{eqnarray}%
\space If we choose $\beta _{2}^{u}=\beta _{3}^{u}=2\beta _{1}^{u}=2\alpha
,\beta _{3}^{d}=2\beta _{2}^{d}=2\beta _{1}^{d}=-2\alpha $, 
the CKM matrix is complex and contains only one phase which is $\alpha $. 
\newline
Using the up and down type quark mass matrices specified in equations (\ref%
{eq:udquarkmatrices}) and their matrix elements provided by equations (\ref%
{eq:mupquarkentries}, \ref{quark2}), we performed a fit of the CKM
observables of the quark sector. 
The scalar rotation matrix $R$ in Eq.  (\ref{quark2}) is 
taken to be a $SO(4)$ rotation matrix
, with the Yukawa couplings on the order of $x\sim z\sim 10^{-1}$. Through this fitting process, we 
estimate the exotic quark masses 
$m_{T_r} $ and the scalar masses $m_{\rho_i} $ 
by selecting a suitable value for the loop function in Eq.  (\ref{eq:1loopfunction}), as follows:
\bea
		m_{T_1} &\approx & 4.0~\text{TeV}, 
		m_{T_2} \approx 5.4~\text{TeV}, 
		m_{\rho_1} \approx 7.0~\text{TeV}, 	\label{mtandrhomasses1}\\
	m_{\rho_2} &\approx& m_{\rho_3} \approx m_{\rho_4}\approx 6.8~\text{TeV}.
	\label{mtandrhomasses2}
\eea

The numerical values in Eqs.(\ref{mtandrhomasses1}), (\ref{mtandrhomasses2}) correspond to a representative
	benchmark points chosen from a scan over the parameter space, and are not meant
	to indicate a precise prediction; physical CP-violating observables are
	insensitive to small variations or rounding of these parameters.

The scalar masses $m_{\rho_i} $ are consistently
greater than the exotic quark masses $m_{T_r} $. This hieararchy can be
explained using the 1-loop function $f(m^2_1,m^2_2)$ as shown in Eq. (\ref%
{eq:1loopfunction}). A lower value of the Yukawa couplings is needed for
higher $m_{\rho_i}$ to balance the entries in the quark mass matries, which
are detailed in Eq. (\ref{eq:udquarkmatrices}). 
To fit the observables in the quark sector, we proceed to vary the quark sector parameters and find the best fit point that minimizes the following $\chi^2$ function:
\begin{align}
\chi ^2 = \sum_{i} \left(\frac{\xi_{obs,i}-\xi_{i}}{\xi_{obs,i}}\right)^2
\end{align}
where $\xi_{obs,i}$ and $\xi_{i}$ are the CKM observables of the quark sector and the
model values of the observables respectively. The obtained values of the
CKM observables are presented in Table (\ref{CKMfix}).

\begin{table}[h!]
\begin{tabular}{|c|c|c|}
\hline
Observable & Model Value & Experimental Value (Ref. \cite%
{ParticleDataGroup:2020ssz}) \\ \hline
$\sin \theta_{12}$ & $0.22518$ & $0.22500 \pm 0.00067$ \\ \hline
$\sin \theta_{23}$ & $0.04179$ & $0.04182^{+0.00085}_{-0.00074}$ \\ \hline
$\sin \theta_{13}$ & 0.00370 & $0.00369 \pm 0.00011$ \\ \hline
$\delta_{CP}$ & $1.148$ & $1.144\pm0.027$ \\ \hline
\end{tabular}%
\caption{The predicted values of the CKM observables. }
\label{CKMfix}
\end{table}
Furthermore, with a variation of a $5\%$ around the best fit values, $%
\delta_{CP}$ varies linearly over $\alpha$ as shown in Fig \ref%
{fig:apha_deltaCP}. 

\begin{figure}[!h]
\centering
\includegraphics[width=0.5\linewidth]{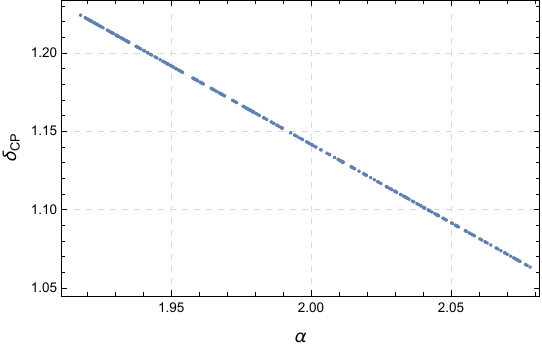}
\caption{$\protect\delta_{CP}$ as a function over $\protect\alpha$ value in
CKM parametrisation.}
\label{fig:apha_deltaCP}
\end{figure}
Our analysis allow a successful determination of the parameters of the CKM quark mixing matrix, with a specific focus on the dynamically generated CP phase
originating from dark-sector interactions. 
The results demonstrate that this framework not only offers a robust description of CP violation but also establishes a direct link between this fundamental asymmetry and new physics beyond the Standard Model. The refined CKM parameters, which include the dark-sector contribution, show remarkable agreement with current experimental data, presenting a compelling alternative to traditional models.

\subsection{Lepton masses and mixings}

In general the SM charged-lepton mass matrix receives tree level and radiative correctiions at different loop order. The SM charged lepton mass matrix  %
can be decomposed 
as follows 
\begin{equation}
M^{(l)} \;=\; M^{(l)(0)} + M^{(l)(1)} + M^{(l)(2)} + \cdots,
\end{equation}
where $M^{(l)(0)}$ denotes the tree-level contribution and $M^{(l)(1)}$ the
one-loop contribution, etc. At tree level the matrix has the rank-one
``third-column'' texture 
\begin{equation}
M^{(l)(0)} \;=\; 
\begin{pmatrix}
0 & 0 & \dfrac{y_1^{(l)}}{\sqrt{2}}v \\[6pt] 
0 & 0 & \dfrac{y_2^{(l)}}{\sqrt{2}}v \\[6pt] 
0 & 0 & \dfrac{y_3^{(l)}}{\sqrt{2}}v%
\end{pmatrix}%
.
\end{equation}
One-loop diagrams (illustrated in Fig.~\ref{charged-lepton}) generate the first- and second-column entries as well as possible complex phases induced by the CP violating dark-sector scalar interactions. Denoting the one-loop corrections by $%
m_{ij}^{(l)}$, the charged-lepton mass matrix up to one-loop order may be
written as 
\begin{equation}
\mathcal{M}^{(l)} \;=\; 
\begin{pmatrix}
e^{-i\alpha} m^{\ell}_{11} & e^{-i\alpha} m^{\ell}_{12} & \dfrac{y^{(l)}_{13}%
}{\sqrt{2}} v \\[6pt] 
e^{-i\alpha} m^{\ell}_{21} & e^{-i\alpha} m^{\ell}_{22} & \dfrac{y^{(l)}_{23}%
}{\sqrt{2}} v \\[6pt] 
e^{-i\alpha} m^{\ell}_{31} & e^{-i\alpha} m^{\ell}_{32} & \dfrac{y^{(l)}_{33}%
}{\sqrt{2}} v%
\end{pmatrix}%
,  \label{eq:leptonmatrix}
\end{equation}
where the overall phase factor $e^{-i\alpha}$ indicates a CP phase communicated from the dark sector to the visible sector; the $m_{ij}^{(l)}$ denote one-loop
induced entries .

Because $M^{(l)(0)}$ is rank one, radiative corrections are necessary to
generate nonzero masses for the electron and muon. The physical mass
eigenvalues and the charged-lepton mixing matrices are obtained by the usual
bi-unitary diagonalization: 
\begin{equation}
V_L^{(l)\,\dagger}\,\mathcal{M}^{(l)}\,V_R^{(l)} \;=\; \func{diag}(m_e,
m_\mu, m_\tau)\,,
\end{equation}
with $V_L^{(l)}$ and $V_R^{(l)}$ the left- and right-handed rotation
matrices for the charged leptons.

\vspace{4pt} \noindent\textbf{Remarks.}

\begin{itemize}
\item The left-handed rotation $V_L^{(l)}$ enters the leptonic mixing (PMNS)
matrix, $U_{\text{PMNS}} = V_L^{(l)\,\dagger} V_L^{(\nu)}$, so the radiative
structure of the charged-lepton mass matrix can affect the observed neutrino
mixing angles and the leptonic Dirac CP violating phase.

\item The magnitude of the one-loop entries $m_{ij}^{(l)}$ is
model-dependent; a typical parametric estimate is 
\begin{eqnarray}
e^{-i\alpha}m_{kn}^{\ell} &=&\dsum\limits_{r=1}^{2}\frac{x_{kr}^{\left(
E\right) }z_{rn}^{\left( l\right) }m_{E_{r}}}{16\pi ^{2}}\left\{
\sum_{i=1}^{4}R_{1i}R_{2i}f\left( m_{\rho _{i}}^{2},m_{E_{r}}^{2}\right)
-\sum_{i=1}^{4}R_{3i}R_{4i}f\left( m_{\rho _{i}}^{2},m_{E_{r}}^{2}\right)
\right\} \\
&&+i\dsum\limits_{r=1}^{2}\frac{x_{kr}^{\left( E\right) }z_{rn}^{\left(
l\right) }m_{E_{r}}}{16\pi ^{2}}\left\{ \sum_{i=1}^{4}R_{1i}R_{4i}f\left(
m_{\rho _{i}}^{2},m_{E_{r}}^{2}\right) +\sum_{i=1}^{4}R_{2i}R_{3i}f\left(
m_{\rho _{i}}^{2},m_{E_{r}}^{2}\right) \right\}.
\end{eqnarray}
\end{itemize}

Therefore, the loop-induced entries in $\mathcal{M}^{(l)}$ are not only
responsible for generating the electron and muon masses, but also serve as a
new source of leptonic mixing and CP violation. This framework naturally
accommodates the observed pattern of neutrino oscillations through the
interplay between the dark-sector phases and the loop corrections in the
charged lepton sector.

\begin{figure}[h]
\centering
\begin{tikzpicture}[x=0.75pt,y=0.75pt,yscale=-1,xscale=1]
 
 	\draw  [draw opacity=0][dash pattern={on 4.5pt off 4.5pt}] (409.6,207.74) .. controls (409.29,165.18) and (444.88,130.39) .. (489.12,130.01) .. controls (533.22,129.63) and (569.3,163.58) .. (569.96,205.94) -- (489.78,207.16) -- cycle ; \draw  [dash pattern={on 4.5pt off 4.5pt}] (409.6,207.74) .. controls (409.29,165.18) and (444.88,130.39) .. (489.12,130.01) .. controls (533.22,129.63) and (569.3,163.58) .. (569.96,205.94) ;  
 	\draw  [dash pattern={on 4.5pt off 4.5pt}]  (490,76) -- (490,136) ;
 	\draw  [dash pattern={on 4.5pt off 4.5pt}]  (489.78,210.16) -- (489.78,270.16) ;
 	\draw    (360,210) -- (410,210) ;
 	\draw [shift={(390,210)}, rotate = 180] [fill={rgb, 255:red, 0; green, 0; blue, 0 }  ][line width=0.08]  [draw opacity=0] (8.93,-4.29) -- (0,0) -- (8.93,4.29) -- cycle    ;
 	\draw    (410,210) -- (489.78,210.16) ;
 	\draw [shift={(443.39,210.07)}, rotate = 0.12] [fill={rgb, 255:red, 0; green, 0; blue, 0 }  ][line width=0.08]  [draw opacity=0] (8.93,-4.29) -- (0,0) -- (8.93,4.29) -- cycle    ;
 	\draw    (490,210) -- (570,210) ;
 	\draw [shift={(535,210)}, rotate = 180] [fill={rgb, 255:red, 0; green, 0; blue, 0 }  ][line width=0.08]  [draw opacity=0] (8.93,-4.29) -- (0,0) -- (8.93,4.29) -- cycle    ;
 	\draw    (570,210) -- (630,210) ;
 	\draw [shift={(593.5,210)}, rotate = 0] [fill={rgb, 255:red, 0; green, 0; blue, 0 }  ][line width=0.08]  [draw opacity=0] (8.93,-4.29) -- (0,0) -- (8.93,4.29) -- cycle    ;
 	
 	\draw (411,132.4) node [anchor=north west][inner sep=0.75pt]    {$\eta $};
 	\draw (501,92.4) node [anchor=north west][inner sep=0.75pt]    {$< \phi ^{*}  >$};
 	\draw (558,132.4) node [anchor=north west][inner sep=0.75pt]    {$\varphi _{2}$};
 	\draw (495,252.4) node [anchor=north west][inner sep=0.75pt]    {$< \sigma^*  >$};
 	\draw (371,222.4) node [anchor=north west][inner sep=0.75pt]    {$l_{iL}$};
 	\draw (591,222.4) node [anchor=north west][inner sep=0.75pt]    {$l_{nR}$};
 	\draw (441,222.4) node [anchor=north west][inner sep=0.75pt]    {$E_{s R}$};
 	\draw (518,222.4) node [anchor=north west][inner sep=0.75pt]    {$E_{kL}$};
 \end{tikzpicture}
\caption{One-loop corrections to the SM charged lepton mass matrices.}
\label{charged-lepton}
\end{figure}

The full neutrino mass matrix in the basis $\left(\nu _{L},\nu
_{R}^{c},N_{R}^{c}\right)$, has the following form: 
\begin{equation}
\mathcal{M}_{\nu }=\left( 
\begin{array}{ccc}
0_{3\times 3} & M_{\nu }^{D} & 0_{3\times 2} \\ 
\left( M_{\nu }^{D}\right) ^{T} & 0_{2\times 2} & M \\ 
0_{2\times 3} & M^{T} & \mu%
\end{array}%
\right) ,
\end{equation}%
where the submatrices, $M_{\nu }^{D},M$, are generated at tree level and are given by:
\begin{equation}
\left( M_{\nu }\right) _{in}=\left( y_{\nu }\right) _{in}\frac{v}{\sqrt{2}},%
\hspace{1cm}\left( M\right) _{nk}=\left( y_{N}\right) _{kn}\frac{v_{\sigma }%
}{\sqrt{2}},
\end{equation}%
with $i=1,2,3$ and $k,n=1,2$. The lepton-number-violating Majorana mass term 
$\mu$ is not generated at tree level; instead, it originates from the
two-loop diagram shown in Fig.~\ref{muterm}. Its structure resembles the one
obtained in Ref.~\cite{Hernandez:2021xet} and can be expressed as 
\begin{eqnarray}
e^{-i \beta}\mu _{sp} &=&\sum_{k=1}^{2}\frac{\left( y_{\Psi }\right)
_{sn}\left( x_{\Omega }^{\ast }\right) _{nk}\left( x_{\Omega }^{\dag
}\right) _{kr}\left( y_{\Psi }^{T}\right) _{rp}m_{\Omega _{k}}}{4(4\pi )^{4}}%
\int_{0}^{1}d\alpha \int_{0}^{1-\alpha }d\beta \frac{1}{\alpha (1-\alpha )}%
\Biggl[I\left( m_{\Omega _{k}}^{2},m_{RR}^{2},m_{RI}^{2}\right) -I\left(
m_{\Omega _{k}}^{2},m_{IR}^{2},m_{II}^{2}\right) \Biggr],  \notag \\
&&s,p,n,k,r=1,2
\end{eqnarray}%
with the loop integral given by \cite{Kajiyama:2013rla}: 
\begin{eqnarray}
I(m_{1}^{2},m_{2}^{2},m_{3}^{2})\!\!\! &=&\!\!\!\frac{m_{1}^{2}m_{2}^{2}\log
\left( \displaystyle\frac{m_{2}^{2}}{m_{1}^{2}}\right)
+m_{2}^{2}m_{3}^{2}\log \left( \displaystyle\frac{m_{3}^{2}}{m_{2}^{2}}%
\right) +m_{3}^{2}m_{1}^{2}\log \left( \displaystyle\frac{m_{1}^{2}}{%
m_{3}^{2}}\right) }{(m_{1}^{2}-m_{2}^{2})(m_{1}^{2}-m_{3}^{2})},  \notag \\
m_{ab}^{2}\!\!\! &=&\!\!\!\frac{\beta m_{\left( \varphi _{1}\right)
_{a}}^{2}+\alpha m_{\left( \varphi _{3}\right) _{b}}^{2}}{\alpha (1-\alpha )}%
\quad (a,b=R:\mathrm{or}:I),
\end{eqnarray}
The small masses of the active neutrinos arise from an inverse seesaw
mechanism at two-loop level and the mass matrices for the physical active
and heavy neutrinos are respectively given by: 
\begin{eqnarray}
\mathcal{M}_{\text{active}}^{\nu } &=&M_{\nu }^{D}\left( M^{T}\right)
^{-1}\mu M^{-1}\left( M_{\nu }^{D}\right) ^{T}, \\
\mathcal{M}_{\text{sterile}}^{\nu (-)} &=&-\frac{1}{2}\left( M+M^{T}\right) +%
\frac{1}{2}\mu , \\
\mathcal{M}_{\text{sterile}}^{\nu (+)} &=&\frac{1}{2}\left( M+M^{T}\right) +%
\frac{1}{2}\mu .
\end{eqnarray}

Owing to the smallness of the $\mu$ parameter, the two pairs of sterile
neutrinos exhibit a tiny mass splitting and thus form pseudo-Dirac states. A
detailed quantitative study of the neutrino sector will be presented
together with the analysis of charged lepton flavor violation in the next
section.

In general, the entries $\mu_{sp}$ are complex, since their loop-induced
origin depends on the interaction couplings of the dark-sector scalar
fields, such as $f_{\varphi_3 \sigma}\,\varphi_3^2 \sigma^{* 2} + \text{h.c.}
$. These terms not only generate the small Majorana mass contributions
required by the inverse-seesaw mechanism, but also act as an additional
source of CP violation in the lepton sector.

Consequently, the model establishes a direct connection between the CP
phases of the charged-lepton and neutrino sectors, linking them to a common
dynamical origin in the dark sector. This interplay provides testable
consequences in neutrino oscillation experiments, thereby offering a direct
probe of the CP-violating dynamics of the dark sector. While this topic
contains many intricate details, we will not delve deeper within the scope
of this work. A more detailed study will be presented in a subsequent
publication. 
\begin{figure}[tbp]
\vspace{-1cm} 
\begin{tikzpicture}[x=0.75pt,y=0.75pt,yscale=-1,xscale=1]
	
	\draw  [draw opacity=0][dash pattern={on 4.5pt off 4.5pt}] (270.1,150.11) .. controls (270.63,106.33) and (306.08,71.03) .. (349.68,71.09) .. controls (393.45,71.14) and (428.91,106.79) .. (429.07,150.8) -- (349.58,151.09) -- cycle ; \draw  [dash pattern={on 4.5pt off 4.5pt}] (270.1,150.11) .. controls (270.63,106.33) and (306.08,71.03) .. (349.68,71.09) .. controls (393.45,71.14) and (428.91,106.79) .. (429.07,150.8) ;  
	\draw  [dash pattern={on 4.5pt off 4.5pt}]  (350,21) -- (350,71) ;
	\draw  [dash pattern={on 4.5pt off 4.5pt}]  (350.13,116.96) -- (350.13,150.96) ;
	\draw    (230,150) -- (270,150) ;
	\draw [shift={(252.6,150)}, rotate = 180] [fill={rgb, 255:red, 0; green, 0; blue, 0 }  ][line width=0.08]  [draw opacity=0] (5.36,-2.57) -- (0,0) -- (5.36,2.57) -- cycle    ;
	\draw  [draw opacity=0][dash pattern={on 4.5pt off 4.5pt}] (390.09,151.14) .. controls (390,175.92) and (372.08,195.96) .. (350.04,195.92) .. controls (328.03,195.87) and (310.21,175.81) .. (310.17,151.07) -- (350.13,150.96) -- cycle ; \draw  [dash pattern={on 4.5pt off 4.5pt}] (390.09,151.14) .. controls (390,175.92) and (372.08,195.96) .. (350.04,195.92) .. controls (328.03,195.87) and (310.21,175.81) .. (310.17,151.07) ;  
	\draw  [dash pattern={on 4.5pt off 4.5pt}]  (350,196) -- (350,230) ;
	\draw    (270,150) -- (310.17,150.07) ;
	\draw [shift={(285.99,150.03)}, rotate = 0.1] [fill={rgb, 255:red, 0; green, 0; blue, 0 }  ][line width=0.08]  [draw opacity=0] (5.36,-2.57) -- (0,0) -- (5.36,2.57) -- cycle    ;
	\draw    (310,150) -- (350.58,150.09) ;
	\draw [shift={(332.89,150.05)}, rotate = 180.13] [fill={rgb, 255:red, 0; green, 0; blue, 0 }  ][line width=0.08]  [draw opacity=0] (5.36,-2.57) -- (0,0) -- (5.36,2.57) -- cycle    ;
	\draw    (349.13,149.96) -- (390,150) ;
	\draw [shift={(365.47,149.98)}, rotate = 0.06] [fill={rgb, 255:red, 0; green, 0; blue, 0 }  ][line width=0.08]  [draw opacity=0] (5.36,-2.57) -- (0,0) -- (5.36,2.57) -- cycle    ;
	\draw    (390,150) -- (429.07,149.8) ;
	\draw [shift={(412.13,149.89)}, rotate = 179.71] [fill={rgb, 255:red, 0; green, 0; blue, 0 }  ][line width=0.08]  [draw opacity=0] (5.36,-2.57) -- (0,0) -- (5.36,2.57) -- cycle    ;
	\draw    (429.07,149.8) -- (470,150) ;
	\draw [shift={(445.43,149.88)}, rotate = 0.28] [fill={rgb, 255:red, 0; green, 0; blue, 0 }  ][line width=0.08]  [draw opacity=0] (5.36,-2.57) -- (0,0) -- (5.36,2.57) -- cycle    ;
	
	\draw (351,13.4) node [anchor=north west][inner sep=0.75pt]  [font=\footnotesize]  {$< \sigma > $};
	\draw (351,113.4) node [anchor=north west][inner sep=0.75pt]  [font=\footnotesize]  {$< \sigma > $};
	\draw (263,81.4) node [anchor=north west][inner sep=0.75pt]  [font=\footnotesize]  {$\varphi _{1\mathnormal{a}}$};
	\draw (421,83.4) node [anchor=north west][inner sep=0.75pt]  [font=\footnotesize]  {$\varphi _{1b}$};
	\draw (351,223.4) node [anchor=north west][inner sep=0.75pt]  [font=\footnotesize]  {$< \sigma > $};
	\draw (239,160.4) node [anchor=north west][inner sep=0.75pt]  [font=\footnotesize]  {$N_{sR}$};
	\draw (280,158.4) node [anchor=north west][inner sep=0.75pt]  [font=\footnotesize]  {$\psi _{nR}^{c}$};
	\draw (319,159.4) node [anchor=north west][inner sep=0.75pt]  [font=\scriptsize]  {${\Om}_{kR}$};
	\draw (361,158.4) node [anchor=north west][inner sep=0.75pt]  [font=\scriptsize]  {${\Om}_{lR}^{\epsilon }$};
	\draw (400,160.4) node [anchor=north west][inner sep=0.75pt]  [font=\footnotesize]  {$\psi _{rR}$};
	\draw (439,161.4) node [anchor=north west][inner sep=0.75pt]  [font=\footnotesize]  {$N_{pR}^{c}$};
	\draw (310,186.4) node [anchor=north west][inner sep=0.75pt]  [font=\footnotesize]  {$\varphi _{3\mathnormal{a}}$};
	\draw (372,187.4) node [anchor=north west][inner sep=0.75pt]  [font=\footnotesize]  {$\varphi _{3b}$};

	\end{tikzpicture}
\caption{Two-loop Feynman diagram contributing to the lepton number violating $\protect\mu $
parameter.}
\label{muterm}
\end{figure}
\newpage
\subsection{Inverse seesaw realization and the Majoron}
\label{subsec:Majorana}

As discussed previously, light neutrino masses in this model are generated via an
inverse seesaw mechanism involving the fields $\nu_L$, $\nu_R^c$, and $N_R^c$.
The explicit structure of the neutrino mass matrix and its diagonalization have
already been presented and will not be repeated here.
General discussions of the inverse seesaw mechanism can be found in
Refs.~\cite{Mohapatra1986,GonzalezGarcia1989,Abada2014}.

The global lepton-number symmetry of the model is spontaneously broken by the
vacuum expectation value of the complex scalar $\sigma$.
As a consequence, the imaginary component of $\sigma$ remains as a physical,
massless Goldstone boson,
\begin{equation}
	\sigma = \frac{1}{\sqrt{2}}\left(v_\sigma + \rho + i J\right),
\end{equation}
which is identified as the Majoron.
The appearance of a Majoron in inverse seesaw realizations with spontaneous
lepton-number breaking is well known
\cite{Chikashige1981,Schechter1982}.

In the present model, the lepton-number-violating parameter $\mu$ is not generated
at tree level, but instead arises radiatively at the two-loop order.
This radiative origin naturally accounts for the smallness of $\mu$ and constitutes
a structural feature of the model.

Since the Majoron originates from the phase of $\sigma$, its couplings to neutrinos
are induced by the same interactions responsible for the generation of $\mu$.
As a result, the effective Majoron--neutrino couplings scale as
\begin{equation}
	g_{J\nu\nu} \sim \frac{\mu}{v_\sigma},
\end{equation}
and vanish in the lepton-number-conserving limit $\mu \to 0$.
This behavior is characteristic of inverse seesaw realizations and differs from
Majoron models in which the couplings are directly proportional to the light
neutrino masses.

\subsection{Phenomenological constraints on the Majoron}
\label{subsec:MajoronPheno}

The phenomenology of Majorons coupled to Majorana neutrinos has been extensively
studied in the literature.
In general, such interactions can induce neutrino decays, contribute to stellar
and supernova energy-loss processes, and affect cosmological observables such as
the effective number of relativistic degrees of freedom $N_{\rm eff}$
\cite{Gelmini1981,Choi1988,Beacom2003,Farzan2008,Raffelt1996}.

In the present model, all Majoron-induced effects involving light neutrinos are
controlled by the effective Majoron--neutrino coupling, which scales as
\begin{equation}
	g_{J\nu\nu} \sim \frac{\mu}{v_\sigma},
\end{equation}
where the lepton-number-violating $\mu$ parameter is radiatively generated at the
two-loop level as shown in the Feynman diagram of Figure \ref{muterm}. This leads to a strong parametric suppression of Majoron interactions.

Astrophysical constraints from stellar cooling and supernova energy loss,
including those derived from SN1987A, require the Majoron--neutrino coupling to
satisfy
\begin{equation}
	g_{J\nu\nu} \lesssim 10^{-5}\text{--}10^{-6},
	\label{gJnunumax}
\end{equation}
in order to avoid excessive energy loss from supernova cores
\cite{Raffelt1996}.
In the parameter region of interest, the couplings
$g_{J\nu\nu} \sim \mu/v_\sigma$ lie well below these bounds. This can be seen considering that in the parameter space of the neutrino sector of the model, discussed in sections \ref{cLFV} and  \ref{sec:diphoton-excess}, we have Max$\left(\mu_{nk}\right)\sim\mathcal{O}\left(10\right)$ keV ($n,k=1,2$), $v_{\sigma}\approx 5$ TeV, which yields very tiny Majoron--neutrino coupling $g_{J\nu\nu}\sim\mathcal{O}\left(10^{-9}\right)$, which is well below the cosmological bound provided by Eq.(\ref{gJnunumax}). As a result, Majoron emission does not significantly affect energy transport in stellar interiors or supernova environments.

Cosmological constraints are likewise satisfied.
The suppressed Majoron--neutrino interaction prevents efficient thermalization
of the Majoron in the early Universe for sufficiently large values of $v_\sigma$,
ensuring consistency with bounds arising from Big Bang nucleosynthesis and cosmic
microwave background observations on $N_{\rm eff}$.
We therefore conclude that the presence of a physical Majoron in the present
model is compatible with existing laboratory, astrophysical, and cosmological
constraints, while remaining directly linked to the origin of neutrino masses
through the radiatively generated lepton-number-violating parameter $\mu$ at two loop level.

The Majoron also couples to the heavy neutral fermions present in the inverse
seesaw sector, since these states acquire mass from Yukawa interactions involving
the scalar field $\sigma$. 

These couplings scale with the corresponding heavy fermion masses and are suppressed by the symmetry-breaking scale $v_\sigma$, but exhibit no direct suppression from the lepton-number-violating parameter $\mu$.

These interactions do not impose new phenomenological constraints within the parameter range considered here. The heavy neutral fermions are too massive to be produced in stellar interiors or supernova cores; consequently, Majoron emission processes involving these heavy states do not contribute to astrophysical energy loss.

\section{Strong CP problem}
\label{strongCP}
Having established a robust framework for weak CP violation, we now extend
our analysis to consider the implications of the dark sector on the strong
CP problem. The existence of a non-zero $\theta$ term in quantum
chromodynamics (QCD) remains a profound puzzle, and our proposed mechanism
offers a novel perspective. By exploring how the same dark-sector
interactions that dynamically generate the weak CP phase might also
influence the strong interaction, we can investigate potential solutions of
the strong CP problem. We first parametrize the up- and down-quark mass
matrices (\ref{eq:udquarkmatrices}) as follows: 
\begin{equation}
M_u = \big( e^{i\alpha} C_u \ , \ r_u \big), \qquad M_d = \big( e^{-i\alpha}
C_d \ , \ r_d \big),
\end{equation}
where $C_{u,d}$ are $3\times 2$ submatrices containing the Yukawa entries
affected by the dark-sector phase, while $r_{u,d}$ denote the third columns,
which are taken real at the considered order. The effective strong CP phase
is defined as 
\begin{equation}
\theta_{QCD}= \text{Arg}\Big[ \text{Det}(M_u)\text{Det}(M_d) \Big].
\end{equation}
Since the first two columns of $M_u$ ($M_d$) carry a common factor $%
e^{i\alpha}$ ($e^{-i\alpha}$), one can factor out $e^{2i\alpha}$ ($%
e^{-2i\alpha}$) from the determinant: 
\begin{equation}
\det(M_u) \propto e^{2i\alpha}, \qquad \det(M_d) \propto e^{-2i\alpha}.
\end{equation}
Multiplying them together gives 
\begin{equation}
\det(M_u)\det(M_d) \;\propto\; e^{2i\alpha} e^{-2i\alpha} = 1,
\end{equation}
and therefore 
\begin{equation}
\theta_{QCD}= 0.
\end{equation}
It means that the strong CP phase vanishes at the one-loop level. The next
step is to examine whether this result remains stable once higher-order
corrections are taken into account. \newline
\begin{figure}[]
\vspace{-1cm} \centering
\includegraphics[width=0.8 \textwidth]{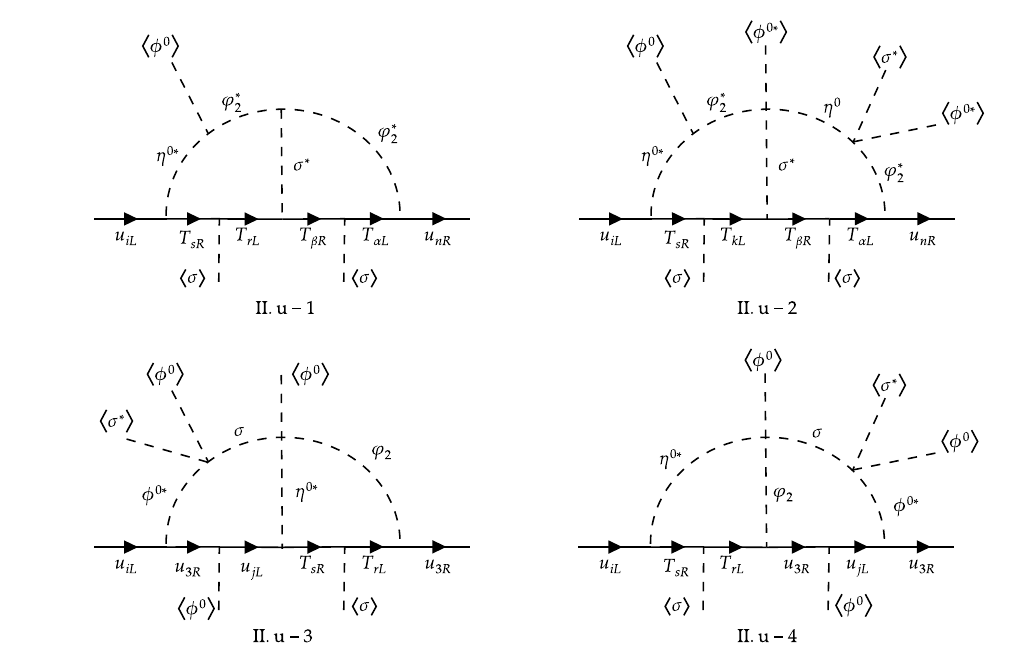}\newline
\includegraphics[width=0.8\textwidth]{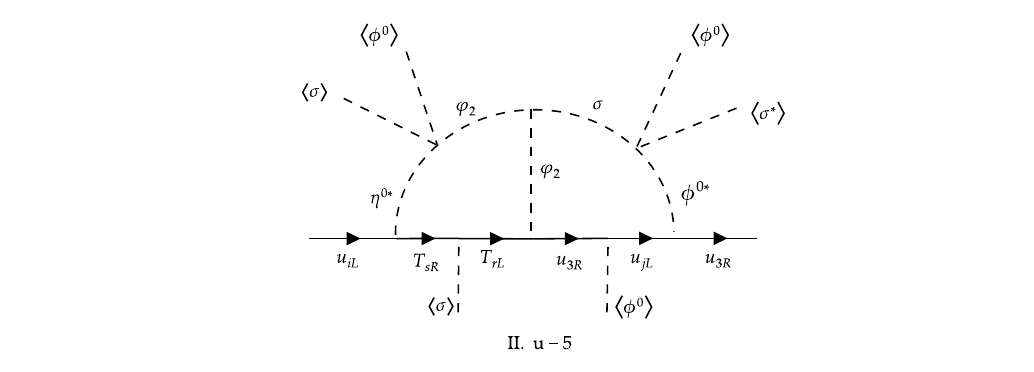} \vspace{0cm}
\caption{ Two-loop corrections to the SM u-quark}
\label{figII-u}
\end{figure}
\begin{figure}[]
\vspace{0cm} \centering
\includegraphics[width=0.8 \textwidth ]{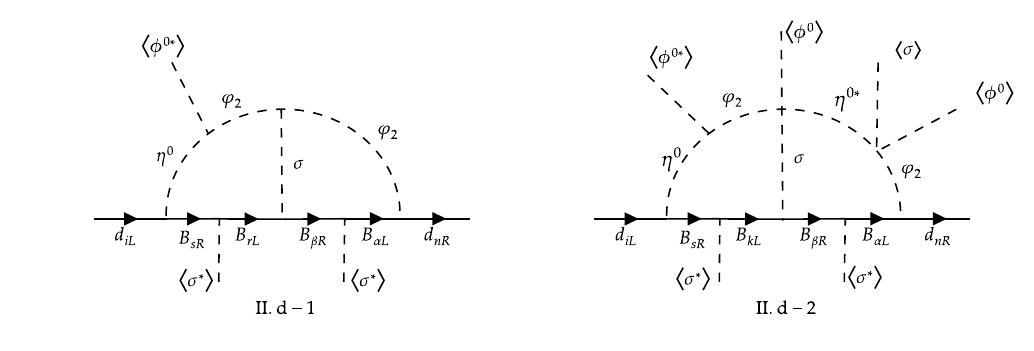}\newline
\includegraphics[width=0.8 \textwidth]{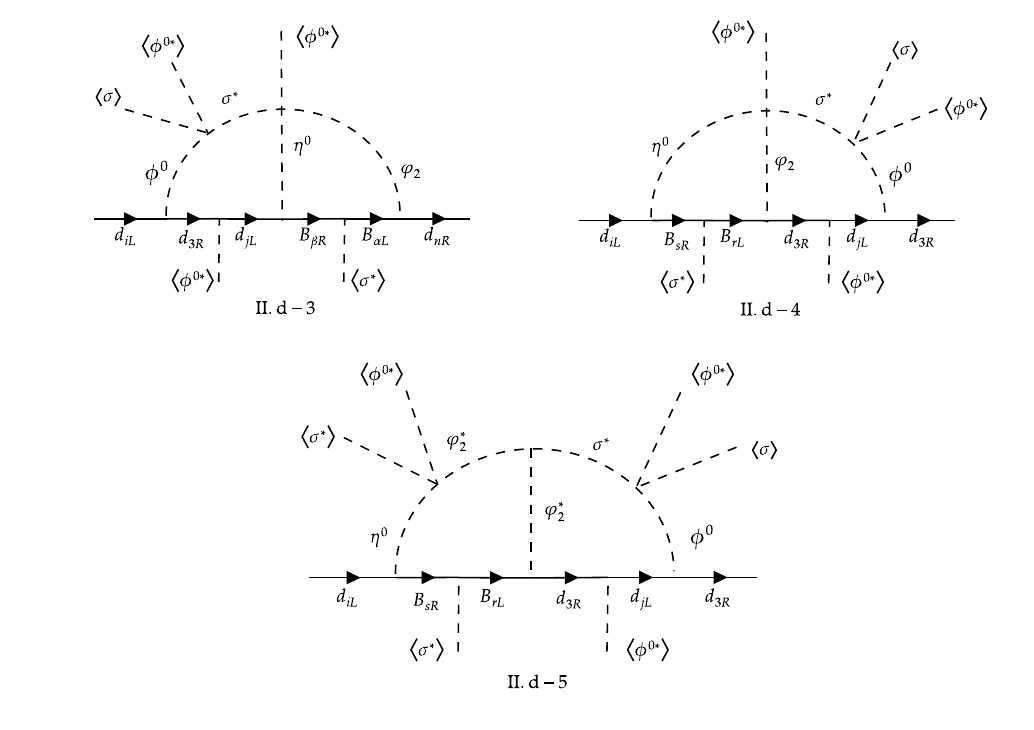} \vspace{0cm}
\caption{ Two-loop corrections to the SM d-quark}
\label{figII-d}
\end{figure}

As displayed in Figures~\ref{figII-u} and ~\ref{figII-d}, the complete set
of two-loop Feynman diagrams contributing to the up- and down-quark mass
matrices exhibit a crucial structural feature: the appearance of complex
phases is uniquely tied to the dark-sector interaction of the form 
\begin{equation}
\mathcal{G}\,\eta^\dagger \phi \varphi_2^\dagger + \text{h.c.}.
\end{equation}
This term induces corrections solely in the first and second columns of the
quark mass matrices, when written in the interaction basis $(u,c,t)$ and $%
(d,s,b)$. As a result, the CP-violating phases are confined entirely to the
light-quark sector, while the two-loop contributions to the third column
remain purely real. This alignment ensures that the determinant phases of
the mass matrices factorize, thereby protecting the strong CP parameter $%
\theta_{QCD}$ from acquiring a nonzero contribution up to two-loop order.

\begin{figure}[H]
	\vspace{0 cm} 
	\centering
	\includegraphics[width=0.7\textwidth]{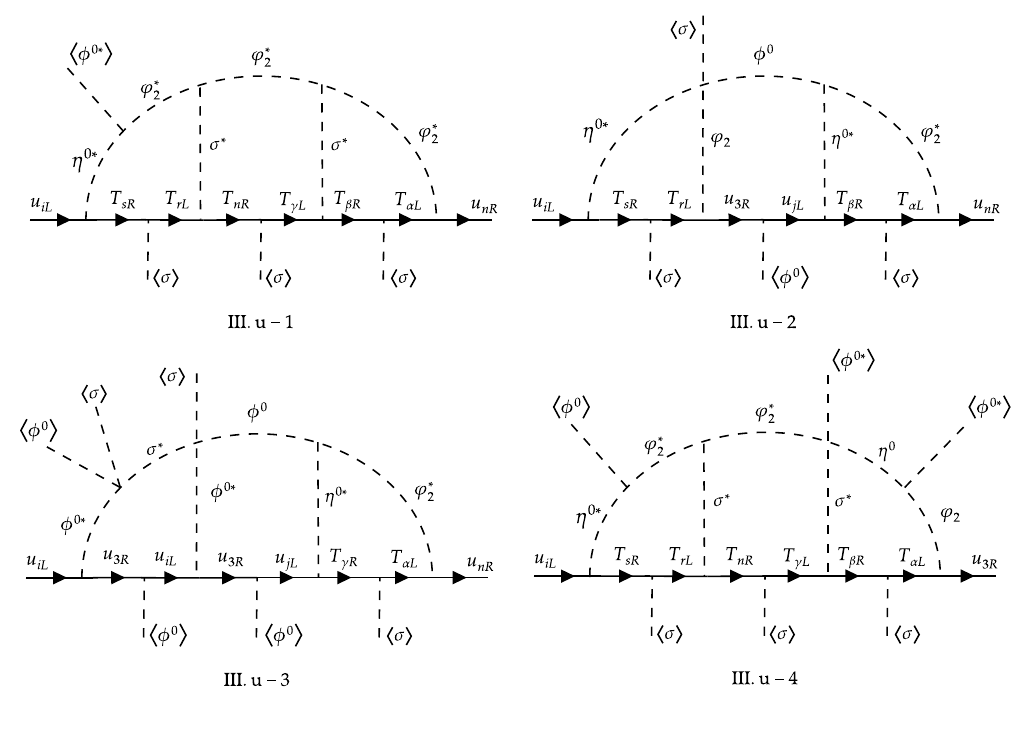} \newline
	\vspace{0.5cm} 
	\includegraphics[width=0.7\textwidth]{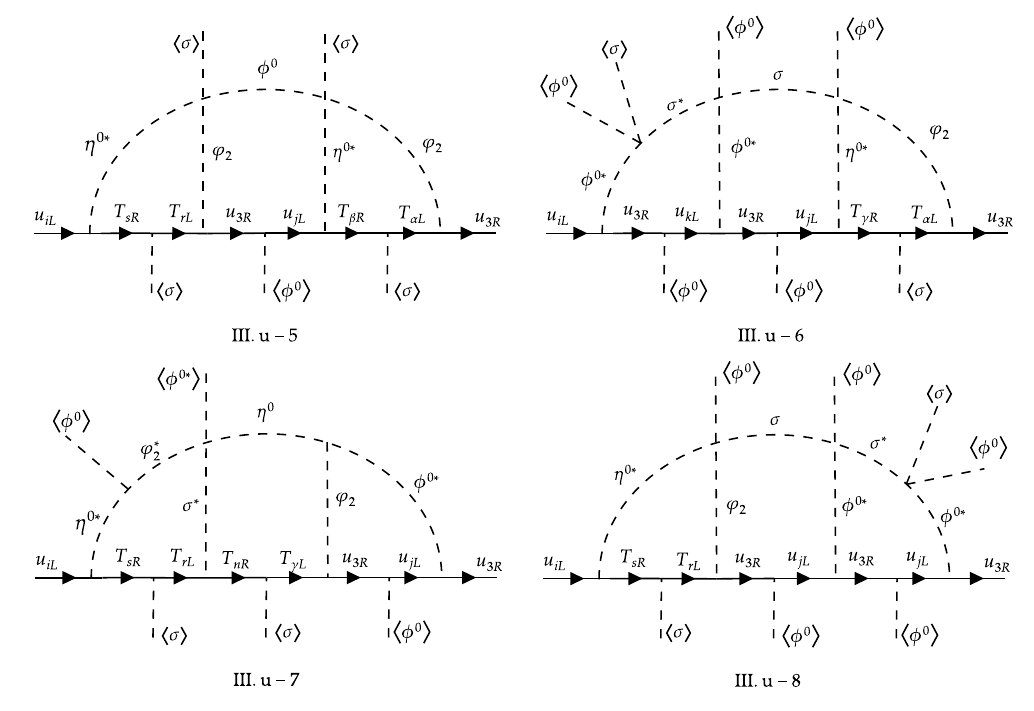} \newline
	\vspace{0.5cm}
	\includegraphics[width=0.7\textwidth]{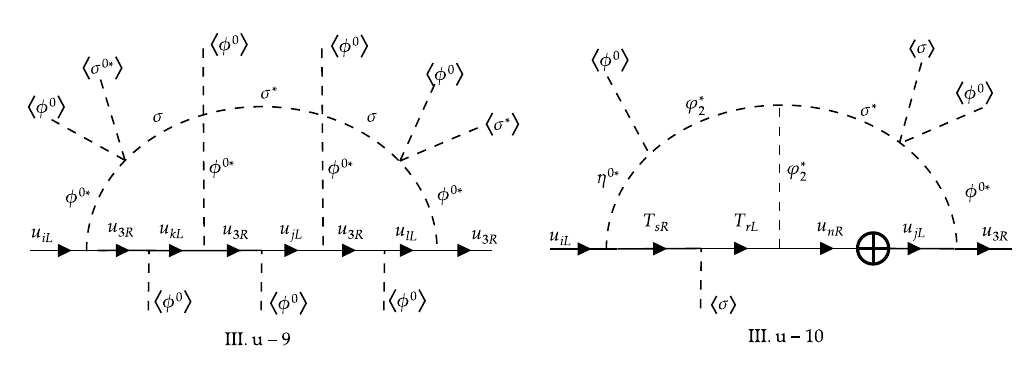} \newline
	\caption{Three-loop corrections to the SM u-quark.}
	\label{figIIIu}
\end{figure}

\begin{figure}[H]
	\centering
	\vspace{0cm}
	\includegraphics[width=0.7\textwidth]{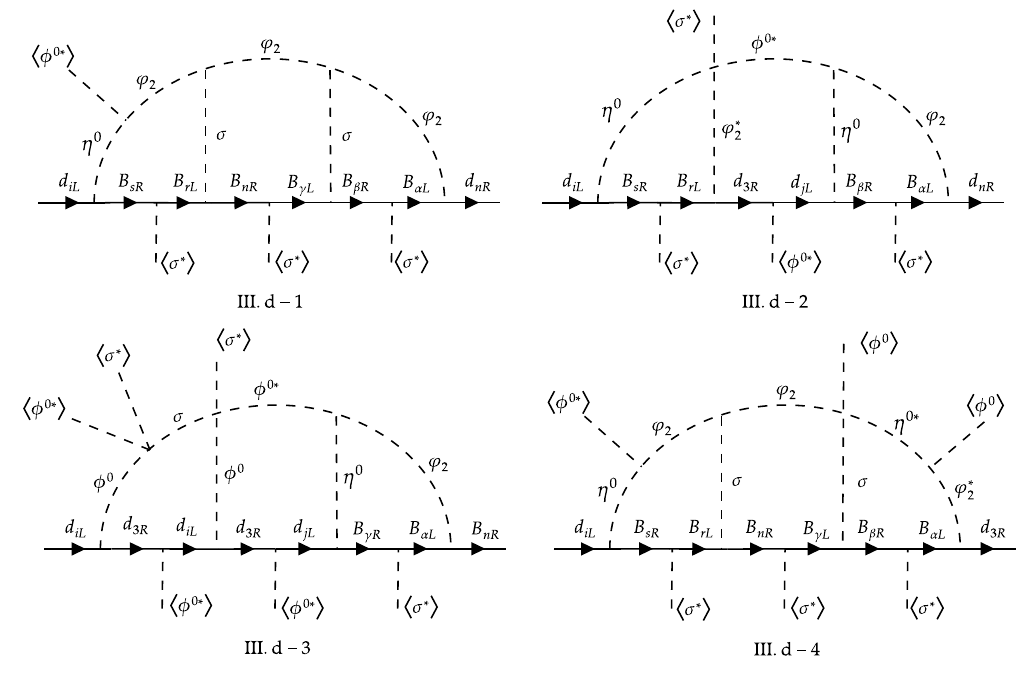} \newline
	\vspace{0.5cm}
	\includegraphics[width=0.7\textwidth]{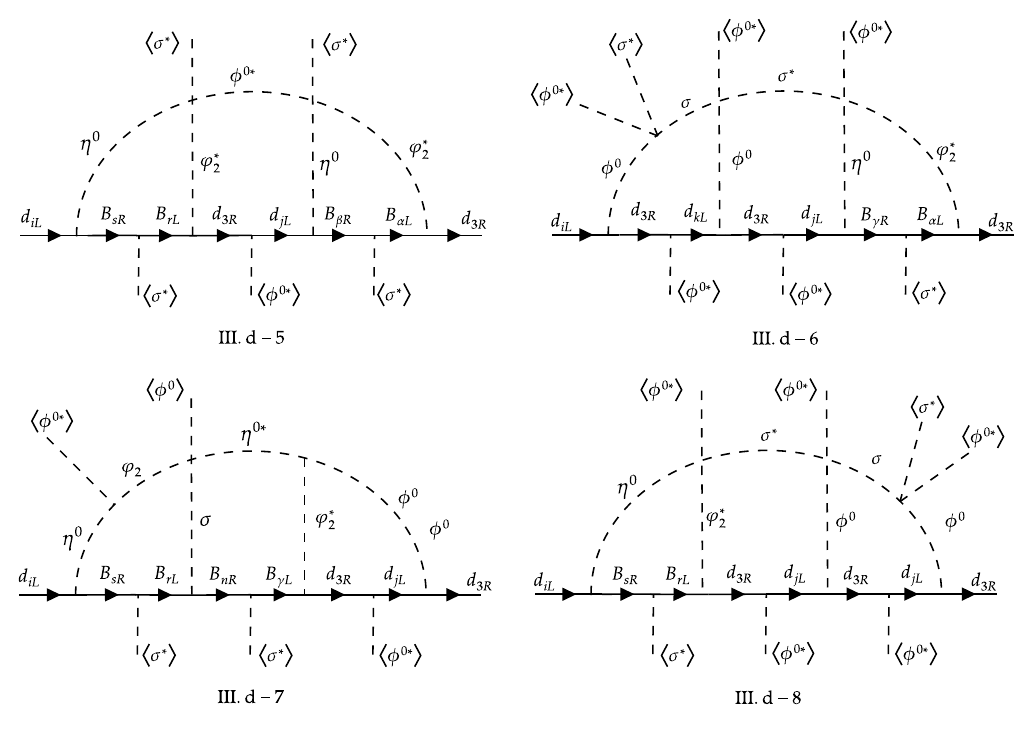} \newline
	\vspace{0.5cm}
	\includegraphics[width=0.7\textwidth]{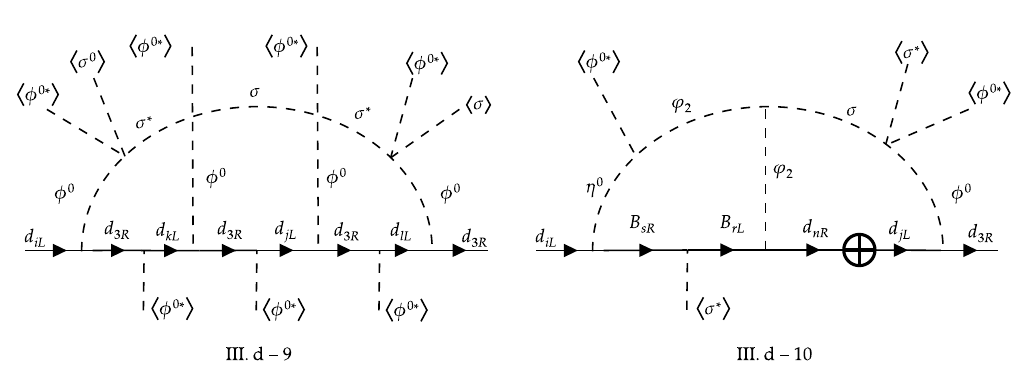}
	\caption{Three-loop corrections to the SM $d$-quark.}
	\label{figIIId}
\end{figure}

We now turn to the next order of radiative corrections. This step is crucial
to determine whether higher-order effects could destabilize the protection
mechanism that prevents the emergence of a nonvanishing strong CP phase. In
particular, we investigate whether three-loop diagrams such as those
depicted in Fig.~(\ref{figIIIu}) and Fig.~(\ref{figIIId}) can reintroduce
complex phases into the quark mass matrices.

The diagrams III.q-1, III.q-2, and III.q-3 with $q=u,d$, shown in Fig.~(\ref%
{figIIIu}) and Fig.~(\ref{figIIId}), contribute to the first two columns of
the quark mass matrices. The complex phases that appear in these columns
trace back to the same dark-sector portal responsible for the one- and
two-loop corrections. As a consequence, the phase pattern of columns 1 and 2
at three loops is identical in structure to that found at lower orders: the
entries in these columns carry a common overall phase (or its conjugate in
the down sector) and therefore continue to factorize in the same manner as
in the one- and two-loop analyses.

With the exception of the two representative topologies denoted III.u-10 and
III.d-10, the three-loop diagrams (III.q-$i$ with $i=4,\dots,9$) that
contribute to the third column of the quark mass matrices exhibit a phase
structure entirely analogous to the lower-order (one- and two-loop)
contributions to the third column. In other words, for the bulk of
three-loop topologies the complex-phase pattern entering $m_{i3}^{q(3)}$
follows the same selection rules and factorization properties that governed $%
m_{i3}^{q(1,2)}$.

The two exceptional diagrams, III.u-10 and III.d-10, at first glance appear
to introduce a new uncompensated phase into the third column. This is
because the candidate diagrams depicted in Fig.~\ref{figIIIu}-III.u-10 and
Fig.~\ref{figIIId}-III.d-10 seem to couple the dark vertex $\mathcal{G}%
\,\eta^\dagger \phi \varphi_2^\dagger + \text{h.c.}$ into a topology that
dresses the third-column operator. However, closing these three-loop
diagrams requires the insertion of a mass sub-diagram on an internal fermion
line. Crucially, that mass insertion is itself generated at one loop and
therefore carries the same dark-sector phase structure (but with the
conjugate phase, once it appears as an insertion closing the fermion line).

Schematically, the three-loop contribution to the mass mixing term for a
representative topology of this type factorizes as 
\begin{equation}
\left(M_{i3}^q \right)^{(3)} \;\propto\; \big(e^{i\alpha}\big)\;\times\;
\left(\big(M^q_{k3}\big)^{{(1)}}\right)^*\;\times\; \mathcal{I}_{3},
\label{eq:3loop-schematic}
\end{equation}
where $\big(M^q_{k3}\big)^{(1)}$ is the one-loop mass insertion (itself $%
\propto e^{i\alpha}$ in the basis where the dark phase appears on columns
1--2), and $\mathcal{I}_{3}$ is the real-valued two-loop integral (after
combining conjugate momentum regions and performing the usual subtractions).

Because the mass insertion appears as a Hermitian term on the fermion
propagator, it contributes with the complex conjugate of the one-loop entry, 
$\big(M^q_{k3}\big)^{(1)}$. Writing 
\begin{equation*}
\left( M^q_{k3}\right)^{(1)}\propto e^{i\alpha}
\left(m_{k3}^{q}\right)^{(1)},
\end{equation*}
the product of phases in Eq.~\eqref{eq:3loop-schematic} becomes 
\begin{equation*}
e^{i\alpha}\times e^{-i\alpha}\,\left(m_{k3}^{q}\right)^{(1)} \;=\;
\left(m_{k3}^{q}\right)^{(1)}.
\end{equation*}

Therefore, although the three-loop topologies IIIu-10 and IIId-10
superficially connect the dark CP vertex to the third-column operator, the
necessary closure via a one-loop mass insertion enforces a conjugation of
the phase and produces an overall real contribution. In other words, the
would-be complex phase from the explicit dark vertex is compensated by the
conjugate phase carried by the mass insertion, so that the net three-loop
correction to $m_{i3}^{q}$ is real. This mechanism explains why the
column-wise factorization of the dark phase persists even after inclusion of
the representative three-loop diagrams, and why no new source of strong CP
violation is introduced at this order.

Taken together, these observations imply that the phase pattern established
at one- and two-loop order persists at three loops: columns 1 and 2 inherit
the same dark-sector-induced phase structure, while the third column remains
real to the order considered. Accordingly, the factorization protection of
the strong CP parameter is maintained at three-loop order unless a
qualitatively different topology or coupling assignment (one that directly
attaches an uncompensated dark phase to the third-column operator) is
introduced.

\section{Radiative CP structure of the quark mass matrices}
\label{subsec:columnwiseCP}

We choose a rephasing basis in which all CP-even Yukawa couplings relevant for
the radiative generation of quark masses are real.
In this basis, the only interaction carrying a physical CP-violating phase and
entering the quark-mass diagrams considered below is
\begin{equation}
	\mathcal{G}\,\eta^\dagger \Phi\,\varphi_2^\dagger + \text{h.c.}
\end{equation}
All other scalar interactions required to close the loops are CP conserving.

Radiative contributions to the first two columns of the up-quark mass matrix,
\begin{equation}
	(M_u)_{in}, \qquad n = 1,2,
\end{equation}
must originate from one of the operators
\begin{equation}
	\bar q_{iL}\tilde{\Phi}\,u_{3R},
	\qquad
	\bar q_{iL}\tilde{\eta}\,T_{kR},
\end{equation}
and terminate on the right-handed external field through
\begin{equation}
	\bar T_{kL}\,\varphi_2^{*}\,u_{nR}.
\end{equation}

The internal fermion lines consist of Dirac mass insertions of the vector-like
quark $T$, possible mixing mass insertions of the form
\begin{equation}
	\bar u_{iL} u_{3R} + \text{h.c.},
\end{equation}
and the CP-even Yukawa interactions
\begin{equation}
	\bar q_{iL}\tilde{\eta}\,T_{kR},
	\qquad
	\bar T_{kL}\varphi_2\,u_{3R},
\end{equation}
all of which are real in the chosen basis.
As a result, the fermion subgraph does not introduce any complex phase, and any
CP violation must originate from the scalar sector.

Since the external field $u_{nR}$ ($n=1,2$) is reached only via $\varphi_2^{*}$,
closing the scalar subgraph requires the presence of the scalar fields
\begin{equation}
	\varphi_2^{*},\ \varphi_2,\ \sigma^{*},\ \eta^{*},\ \Phi,
\end{equation}
arranged such that all scalar lines can be consistently contracted.
Under these assumptions, exactly one insertion of the CP-violating interaction
\begin{equation}
	\mathcal{G}\,\eta^\dagger \Phi\,\varphi_2^\dagger
\end{equation}
is required, while any additional loop insertions involve only CP-conserving
vertices.
Consequently, higher-loop corrections do not generate new independent phases and
preserve the same column-wise phase structure as at lower orders.
An analogous argument applies to the down-quark mass matrix.

For the third column, the radiative contributions again originate from
\begin{equation}
	\bar q_{iL}\tilde{\Phi}\,u_{3R},
	\qquad
	\bar q_{iL}\tilde{\eta}\,T_{kR},
\end{equation}
but terminate on $u_{3R}$ through
\begin{equation}
	\bar T_{kL}\,\varphi_2\,u_{3R}.
\end{equation}
In this case, the scalar subgraph can be closed entirely by CP-conserving scalar
interactions, leading to a real contribution.
If the CP-violating interaction $\eta^\dagger \Phi \varphi_2^\dagger$ is inserted,
closure of the loop requires the simultaneous insertion of its hermitian
conjugate $\Phi^\dagger \eta \varphi_2$.
Thus, CP-violating effects enter the third-column amplitudes only through the
CP-even combination $|\mathcal{G}|^2$, and no complex phase is generated.

The resulting structure of the quark mass matrices can therefore be written as
\begin{equation}
	M_u = \bigl(e^{i\alpha} C_u,\ r_u\bigr),
	\qquad
	M_d = \bigl(e^{-i\alpha} C_d,\ r_d\bigr),
\end{equation}
where $C_{u,d}$ and $r_{u,d}$ are real in the chosen basis.
As long as this structure is preserved, the rephasing-invariant combination
\begin{equation}
	\bar{\theta} = \arg \det(M_u M_d)
\end{equation}
vanishes identically within this class of radiative contributions.

\section{Dark matter phenomenology}

\label{dm}

Understanding the spectrum and stability of dark-sector states is essential
to connect the model with cosmological observables and terrestrial searches.
After spontaneous symmetry breaking, the residual discrete symmetry is $%
\mathbb{Z}_2 \times \mathbb{Z}_2^{\prime }$. Such remnant discrete symmetry $%
\mathbb{Z}_2 \times \mathbb{Z}_2^{\prime }$ is crucial for ensuring the stability of the scalar and/or fermionic dark matter candidates. Each dark matter candidate 
carries a pair of
charges $(q,q^{\prime })$ with $q,q^{\prime }\in\{0,1\}$, which naturally
classifies the dark-sector states into three distinct categories, 
\begin{equation*}
(q,q^{\prime }) \in \{(0,1),(1,0),(1,1)\},
\end{equation*}
hereafter denoted as $\mathrm{DM}_1,\mathrm{DM}_2,\mathrm{DM}_3$ (see Table~%
\ref{tableDM}).

\begin{table}[H]
\centering
\begin{tabular}{|c|c|}
\hline
$(Z_2, Z_2^\prime )$ & Representative fields \\ \hline
$\mathrm{DM}_1 \equiv (1,0)$ & $\eta$, $\varphi_2$, $\Omega_{nR}$ \\ \hline
$\mathrm{DM}_2 \equiv (0,1)$ & $\varphi_3$ \\ \hline
$\mathrm{DM}_3 \equiv (1,1)$ & $\varphi_1$, $\Psi_{nR}$ \\ \hline
\end{tabular}%
\caption{Classification of dark-sector fields according to their $%
(Z_{2},Z_{2}^{\prime })$ charges.}
\label{tableDM}
\end{table}

The stability pattern and mass ordering determine whether the model realizes
a two- or three-component dark-matter scenario. In particular, if $\mathrm{DM%
}_1$ and $\mathrm{DM}_2$ are the lightest states and 
\begin{equation*}
m_{\mathrm{DM}_3} > m_{\mathrm{DM}_1}+m_{\mathrm{DM}_2},
\end{equation*}
then $\mathrm{DM}_3$ efficiently decays into a $\mathrm{DM}_1+\mathrm{DM}_2$
pair, resulting in an effective two-component dark sector. Otherwise, all
three species can contribute to the relic abundance.

In the following, we concentrate on the two-component scenario, which can be
realised in two ways:

\begin{itemize}
\item Scalar-scalar system: both dark components are scalars,
namely $\varphi_3$ and a mixed state $\rho_i$ arising from $\eta$ and $%
\varphi_2$;

\item Scalar-fermion system: one dark component is the scalar $%
\varphi_3$, while the other one is the fermion $\Omega_{nR}$.
\end{itemize}

The relic abundances of these dark components are determined by the standard
freeze-out mechanism. Rather than presenting the explicit coupled Boltzmann
equations, we employ the public package \texttt{micrOMEGAs}~6.2.3 package~%
\cite{Alguero:2023zol}, which incorporates all annihilation, coannihilation
and conversion processes and provides numerical predictions for the thermal
relic density. The code also computes elastic scattering cross sections
relevant for direct detection, as well as annihilation rates relevant for
indirect searches.

In order to extract quantitative predictions, we first identify
representative benchmark points in the parameter space of the model. The
choice is guided by following considerations:

\begin{enumerate}
\item Theoretical consistency: perturbativity of the couplings, vacuum
stability of the scalar potential are imposed to
ensure the validity of the effective description.

\item Collider limits: the masses of new gauge bosons and scalars are chosen
above the current LHC bounds \cite{ParticleDataGroup:2024cfk}, while mixing angles between new states and the
SM Higgs sector are kept within the limits set by Higgs precision data \cite{ATLAS:2022vkf,CMS:2022dwd,ParticleDataGroup:2024cfk}.

\item 
Dark-sector spectrum: The mass hierarchy in the dark sector is arranged such that $\mathrm{DM}_1$ and $\mathrm{DM}_2$ constitute the stable relics, providing a concrete realization of the two-component dark matter scenario discussed above.
\end{enumerate}

Concretely, we scan over the following ranges of parameters:

\begin{itemize}
\item \textbf{Self-couplings.} The SM Higgs quartic $\lambda_\phi$ is fixed
by the observed $125$~GeV Higgs mass, $m_h^2 \simeq \lambda_\phi v^2$,
giving $\lambda_\phi \sim \frac{1}{2}$. For simplicity, we assume $%
\lambda_{\sigma,\eta,\varphi_i}$ are of the same order and sample them in
the range $[0,1]$.

\item \textbf{Mixed couplings.} The portal couplings $\lambda_{ij}$ $(i\neq
j)$ may be either comparable to the self--couplings ($\lambda_{ij}\sim%
\lambda_i$) or smaller ($\lambda_{ij}\sim0.1\,\lambda_i$). The same
assumptions are applied to the couplings in the $V_{\text{CPV}}, V^\prime_{%
\text{CPV}}$, the real couplings 
as well as the real and imaginary parts of $\mathcal{G},f^{\prime
},f_\sigma,f^{\prime }_\sigma$. In practice, we randomly sample these
couplings in two ranges: $[-0.1,0.1]$ and $[-0.01,0.01]$.

\item \textbf{Dark matter masses.} The masses of the two DM components,
either $(m_{\phi_3},m_{\rho_1})$ or $(m_{\phi_3},m_{\Omega_{nR}})$, are
varied in the range $[200,3000]$~GeV. For the heavier dark state DM$_3$, we
set $m_{\Psi_{nR}}=5000$~GeV and $m_{\phi_1}=5500$~GeV, such that $m_{\text{%
DM}_1}+m_{\text{DM}_2}<m_{\text{DM}_3}$.

\item \textbf{VEV.} The vacuum expectation value $v_\sigma$ is scanned
within $[100,3000]$~GeV.
\end{itemize}

\begin{figure}[]
\vspace{0cm} {\includegraphics[width=12cm,height=5.0cm]{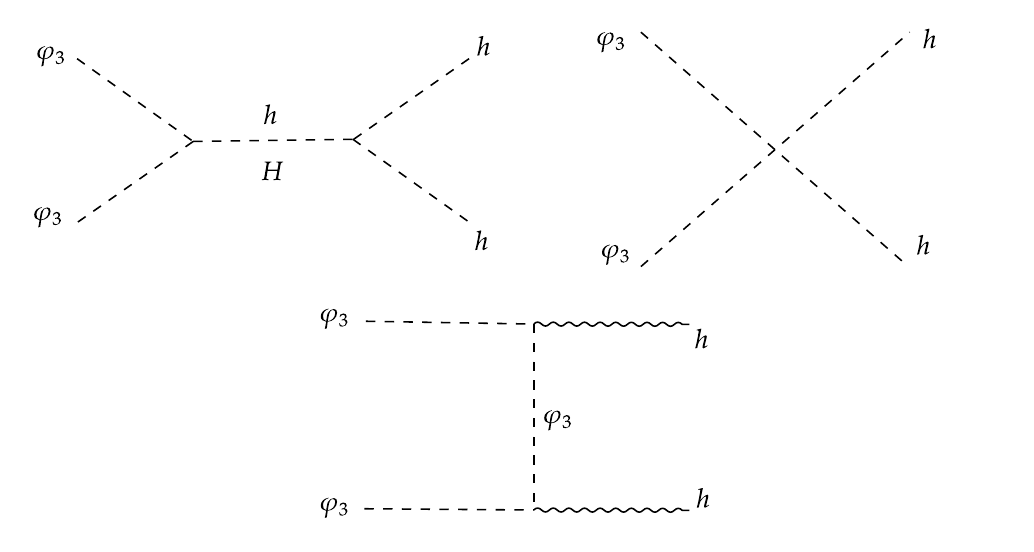}}
\caption{Annihilation of the DM, $\protect\varphi_3$, via a SM-like
Higgs/new Higgs portals}
\label{ScalarDM2}
\end{figure}

\begin{table}[tbp]
\begin{tabular}{|c|c|}
\hline
\text{Parameters} & \text{Ranges} \\ \hline
$\lambda_{\phi,\eta,\sigma,\varphi_{1,2,3}}$ & $[0,0.1]$ \\ \hline
$\lambda_{ij} (i \neq j)$ & $[-0.1,0.1]$ \text{or} $[-0.01,0.01]$ \\ \hline
$v_{\sigma}$ & $[100,3000]$ GeV \\ \hline
$m_{\phi_3,\rho_1}$ & $[200,3000]$ GeV \\ \hline
$m_{\phi_3,\Omega_{1R}}$ & $[200,3000]$ GeV \\ \hline
\end{tabular}
\caption{Input parameters ranges for numerically study of DM}
\label{inputDM}
\end{table}
First, we assume that both DM components are scalar fields, $\varphi _{3}$
and $\rho _{1}$, produced via the freeze-out mechanism. For the sake of
simplicity, we consider a scenario where the scalar field $\varphi _{3}$ is
mainly a weak singlet component whereas $\rho _{1}$ is mainly composed of
the neutral CP even component of a $SU\left( 2\right) _{L}$ electroweak
doublet. Hence, $\varphi _{3}$ annihilates to normal matter only via the
Higgs portal, $\rho _{1}$, whereas $\varphi _{3}$ annihilates to the SM
fields via both the Higgs and gauge portals, as shown in Figs (\ref%
{ScalarDM1},\ref{ScalarDM2}). \newline
\begin{figure}[]
\begin{center}
\vspace{0cm}{\includegraphics[width=14cm,height=15.0cm]{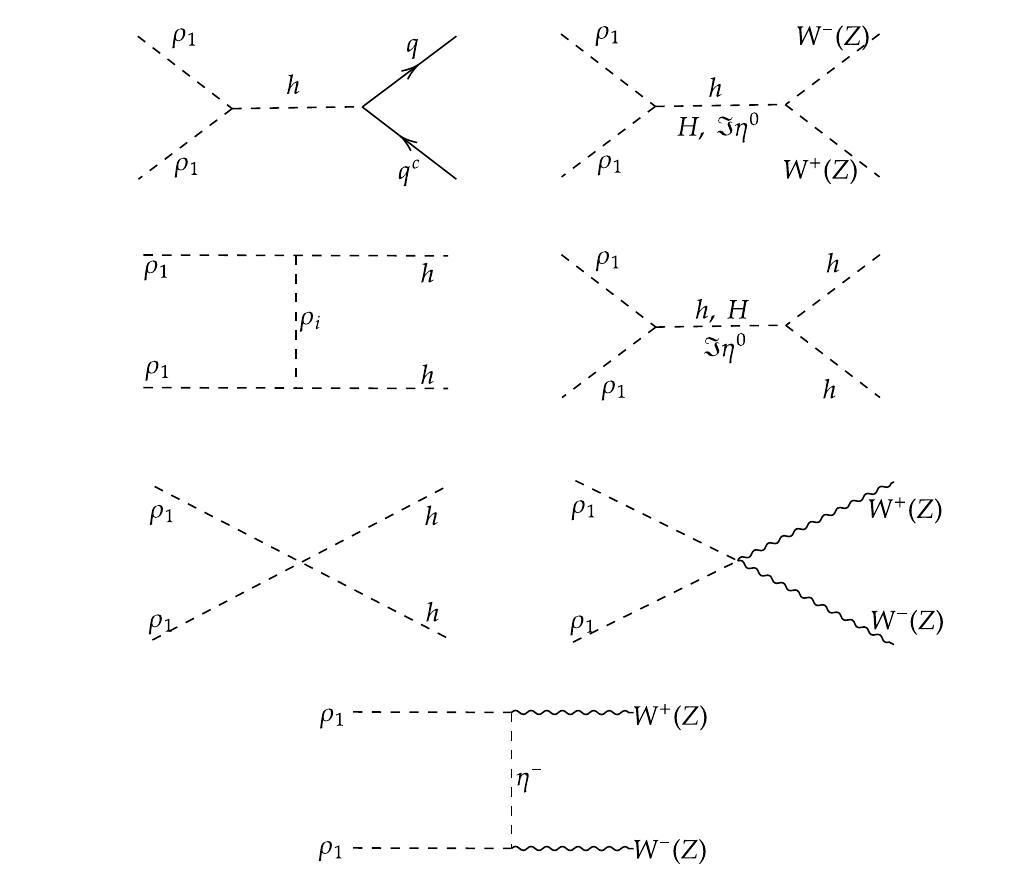}}
\end{center}
\caption{Annihilation of the DM, $\protect\rho _{1}$, via a SM-like
Higgs/new Higgs and gauge bosons portals}
\label{ScalarDM1}
\end{figure}

\begin{figure}[]
{
	\includegraphics[width=10cm,height=4cm]{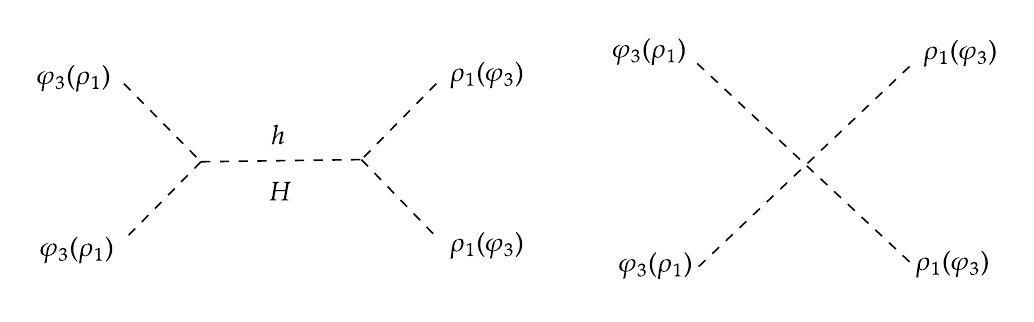}}\newline
\vspace{0cm} {\ }
\caption{Conversion between two-component scalar DM.}
\label{ScalarDM3}
\end{figure}

\begin{figure}[]
\begin{center}
\vspace{0cm} 
{
\includegraphics[width=8cm,height=6.5cm]{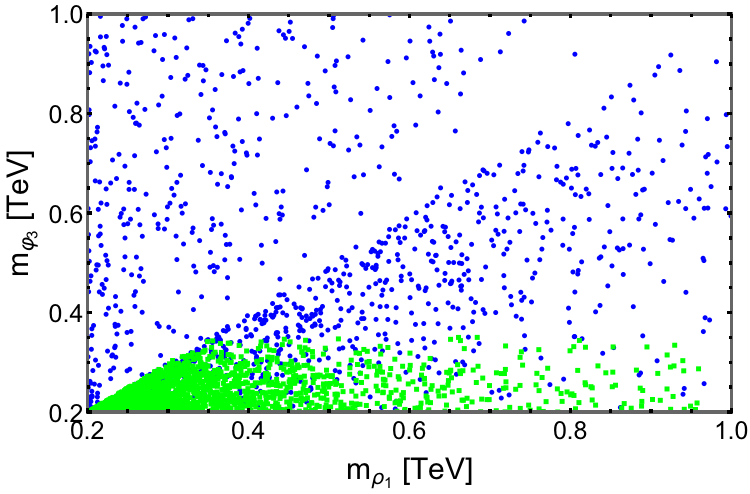}} 
\vspace{-0cm} {\ }
\end{center}
\caption{Relation between the masses of two scalar dark matter components 
$m_{\protect\rho_1}$ and $m_{\protect\varphi_3}$ satisfying $3\protect\sigma$
experimental constraint of DM relic density $\Omega_{\text{DM}%
}h^2=0.11933\pm0.00091$ \protect\cite{Planck:2018vyg}. The blue and green
points are plotted for two cases of mixed couplings $\protect\lambda_{ij}\in
[-0.1,0.1]$ and $[-0.01,0.01]$, respectively.}
\label{DMrelic1}
\end{figure}

Furthermore, there is a conversion between two types of DM, where the
heavier DM component can annihilate into the lighter one. We introduce an
additional annihilation process. 
\begin{equation}
\rho _{1}\rho _{1}\rightarrow \varphi _{3}\varphi _{3}\quad \quad if\quad
m_{\rho _{1}}>m_{\varphi _{3}},
\end{equation}%
or 
\begin{equation}
\varphi _{3}\varphi _{3}\rightarrow \rho _{1}\rho _{1}\quad \quad if\quad
m_{\varphi _{3}}>m_{\rho _{1}},
\end{equation}%
The Feynman diagrams for conversion between two-component scalar DM are
presented in Fig. (\ref{ScalarDM3}). 
We establish the correlation between the masses of the two scalar DM
components fulfilling DM relic abundances \cite{Planck:2018vyg}, as
illustrated in Fig.(\ref{DMrelic1}). The contributions of both scalar DM
candidates depend on their respective masses. The viable mass range for dark
matter compatible with the relic density extends up to the TeV scale.\newline

As a result, the steep dependence of the total dark matter density on the
mixing parameter is primarily governed by the singlet contribution, while
the doublet provides a more robust and less sensitive background. In
particular, the constraint on the mass of the singlet scalar component $%
\varphi_3$ depends strongly on the range of the mixed couplings $%
\lambda_{ij} $. For small values, $\lambda_{ij}\in[-0.01,0.01]$, the singlet
mass is tightly bounded as $m_{\varphi_3}\lesssim 400~\text{GeV}$ in order
to reproduce the correct relic density. In contrast, for larger couplings, $%
\lambda_{ij}\in[-0.1,0.1]$, the annihilation of $\varphi_3$ is sufficiently
enhanced and its mass can reach the TeV scale without overclosing the
Universe.

On the other hand, the relic density of the doublet Higgs component $\rho_1$
is mainly governed by gauge-mediated annihilation channels into electroweak
gauge bosons such as $W^+W^-$ and $ZZ$. These processes remain efficient
even for small Higgs mixing, so the allowed mass region of $\rho_1$ is only
mildly affected by the choice of $\lambda_{ij}$ and can consistently extend
to the TeV scale in both coupling scenarios.

This distinction between singlet and doublet behavior illustrates the
different interaction structures: while the singlet $\varphi_3$ is
dominantly probed through Higgs-portal direct detection, the doublet $\rho_1$
is more naturally constrained by gauge-mediated annihilation channels and
can be effectively tested in indirect and collider searches.

\begin{figure}[h!]
	\centering
	\subfigure[]{%
		\includegraphics[width=7.8cm,height=6.2cm]{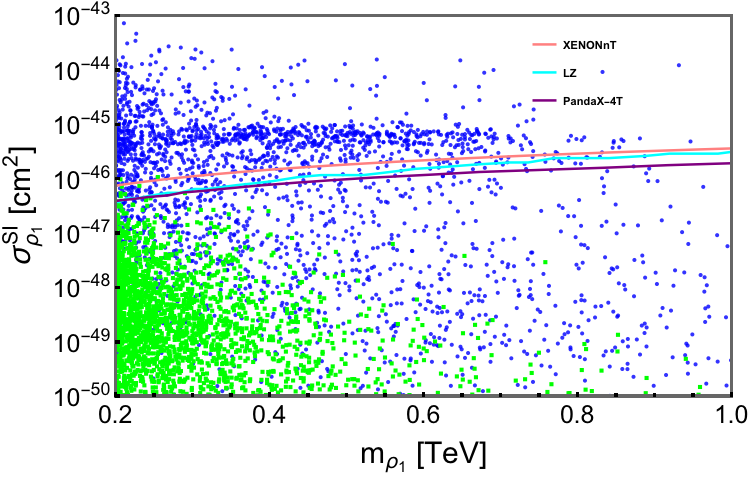}%
	} \hspace{0.6cm}
	\subfigure[]{%
		\includegraphics[width=7.8cm,height=6.2cm]{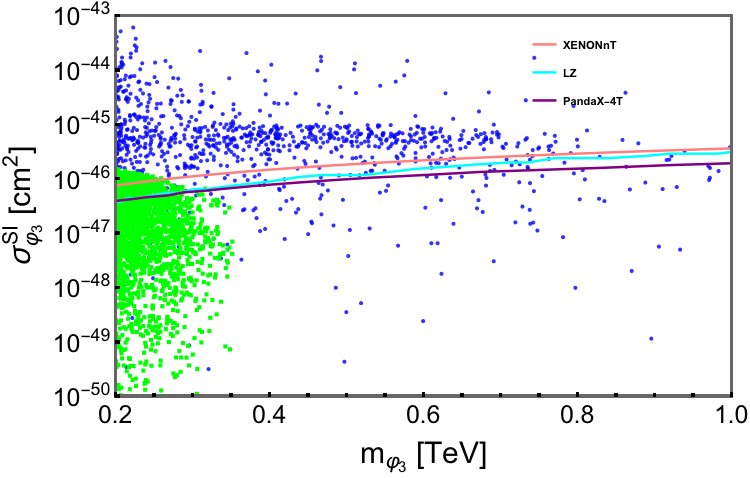}%
	}
	\vspace{0.2cm}
	\caption{The left and right panels demonstrate the dependence of the
		spin-independent scattering cross-section of each DM component 
		$\rho_{1}$ and $\varphi_{3}$ with nucleons on the DM masses 
		$m_{\varphi_3}$ and $m_{\rho_1}$, respectively. 
		The blue and green points correspond to two cases of mixed couplings 
		$\lambda_{ij}\in[-0.1,0.1]$ and $[-0.01,0.01]$, respectively. 
		The pink, cyan, and purple solid lines represent the current upper experimental limits reported by 
		XENONnT~\cite{XENON:2025vwd}, LZ~\cite{LZ:2022lsv}, and PandaX-4T~\cite{PandaX:2024qfu}, respectively.}
	\label{DD_2S}
\end{figure}
Having clarified the distinct relic density dependence of the singlet and
doublet components, we now turn to the implications for direct detection.
Imposing the $3\sigma$ constraint on the relic density~\cite{Planck:2018vyg}%
, we obtain the left and right panels of Fig.~\ref{DD_2S}, which display the
spin-independent (SI) scattering cross section of $\rho_{1}$ and $%
\varphi_{3} $ with nucleons as a function of their masses $m_{\rho_1}$ and $%
m_{\varphi_3} $. Two representative scenarios of mixed couplings are
considered: $\lambda_{ij}\in[-0.1,0.1]$ (blue points) and $\lambda_{ij}\in[%
-0.01,0.01]$ (green points).

We observe that for relatively large mixing, $\lambda_{ij}\in[-0.1,0.1]$,
both $\rho_1$ and $\varphi_3$ exhibit sizable SI cross sections, with the
majority of $\varphi_3$ points lying above the current exclusion limits. In
contrast, for smaller mixing values, $\lambda_{ij}\in[-0.01,0.01]$, most
parameter points fall below the experimental upper bounds, thus remaining
viable. This trend directly reflects the quadratic dependence of the SI
scattering rate on the mixed couplings, $\sigma^{\text{SI}}_{\text{DM}%
}\propto \lambda_{ij}^2$, so that suppressing $\lambda_{ij}$ naturally
reduces the cross section.

These results reinforce the complementarity between relic density and direct
detection constraints: while the singlet scalar $\varphi_3$ is strongly
probed through Higgs-portal interactions and subject to stringent direct
detection limits, the doublet $\rho_1$ benefits from gauge-mediated
annihilation channels and is correspondingly less constrained, even in the
large-mixing regime.

\begin{figure}[H]
\centering
\vspace{0cm} {\includegraphics[width=12cm,height=4.5cm]{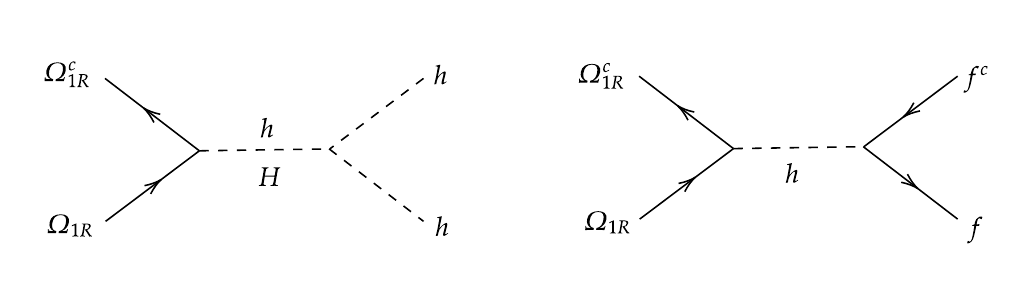}} \newline
\vspace{0cm}{\ }
\caption{Annihilation of fermion DM $\Omega _{nR}$. }
\label{FermionDM1}
\end{figure}
\begin{figure}[h!]
	\centering
	\includegraphics[width=12cm,height=5cm]{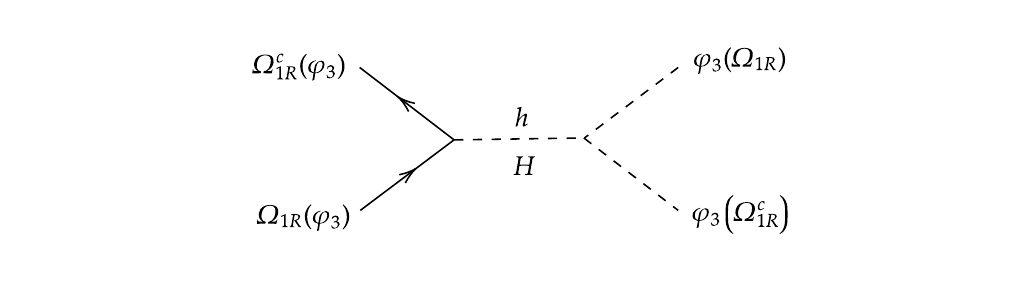}
	\caption{Conversion between fermion DM $\Omega_{nR}$ and scalar DM $\varphi_{3}$.}
	\label{FermionDM2}
\end{figure}

Next, we analyze a scenario involving a fermion and a scalar DM candidate.
In this case, we consider $\Omega_{1R}$ and $\varphi_{3}$ as two-component
DM fields. The annihilation process of the fermionic field, $\Omega_{1R}$,
into SM fields is illustrated by the diagrams in Fig. (\ref{FermionDM1}). 
The conversion between fermion and scalar DM components is presented in Fig.(%
\ref{FermionDM2}), which assumes that $m_{\Omega _{1R}}>m_{\varphi _{3}}$. 

\begin{figure}[]
\begin{center}
\vspace{-0.8cm} 
{\includegraphics[width=10cm,height=7.0cm]{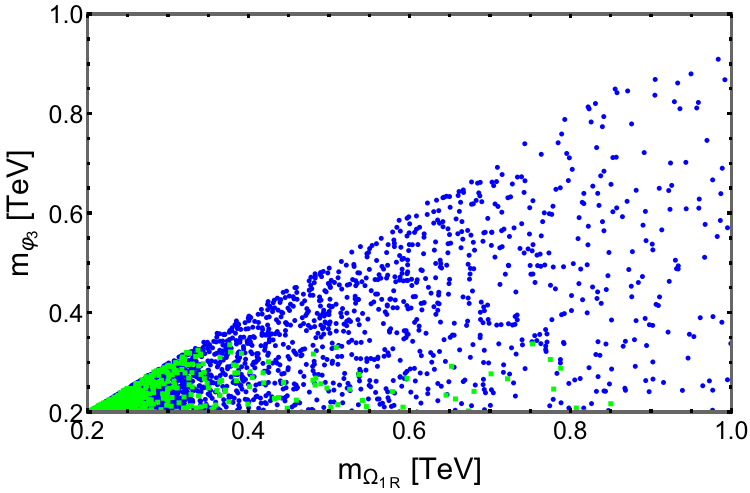}} 
\vspace{0.2 cm} {\ }
\end{center}
\caption{The figure demonstrates the relation between the masses of two
DM components $m_{\Omega_{1R}}$ and $m_{\protect\varphi_3}$ satisfying $3%
\protect\sigma$ experimental constraint of DM relic density $\Omega_{\text{DM%
}}h^2=0.11933\pm0.00091$ \protect\cite{Planck:2018vyg}. The blue and green
points are plotted for two cases mixed couplings $\protect\lambda_{ij}\in
[-0.1,0.1]$ and $[-0.01,0.01]$, respectively.}relation
\label{DM_relic_SF}
\end{figure}

\begin{figure}[h]
	\centering
	\vspace{0cm}
	\subfigure[]{%
		\includegraphics[width=7.8cm,height=6.2cm]{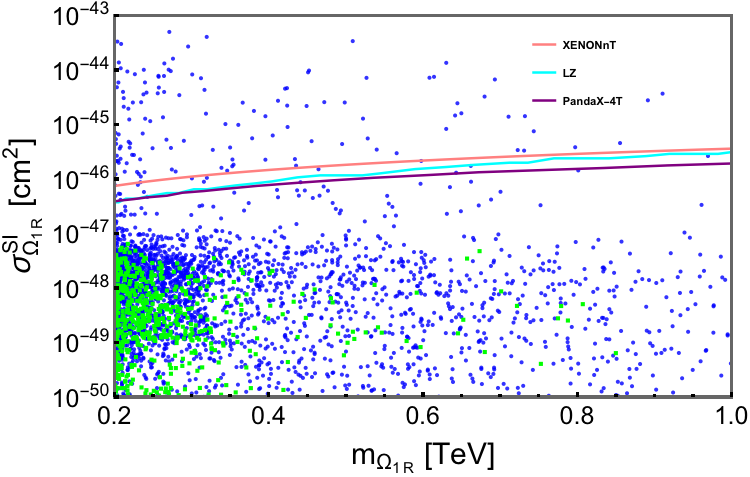}%
	} \hspace{0.5cm}
	\subfigure[]{%
		\includegraphics[width=7.8cm,height=6.2cm]{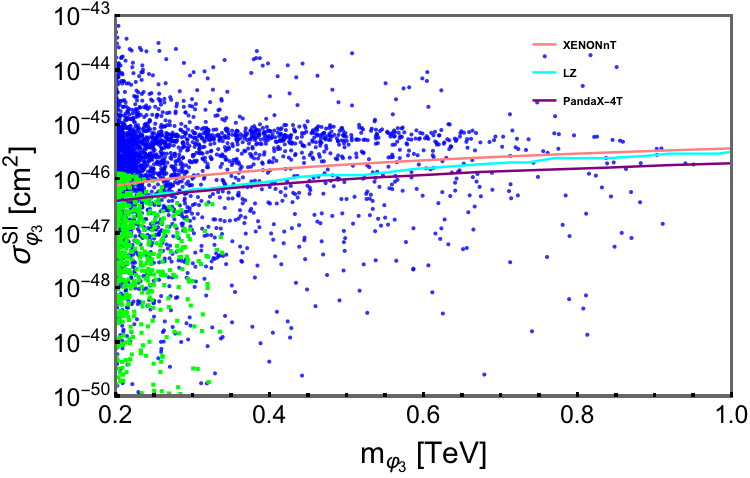}%
	}
	\vspace{0.2cm}
	\caption{The left and right panels demonstrate the dependence of
		spin-independent of each DM component $\Omega _{1R}$, $\varphi _{3}$%
		-nucleon scattering cross-section on DM masses $m_{\Omega_{1R}}$ and $m_{%
			\rho_1}$, respectively. The blue and green points are for two cases
		of mixed couplings $\lambda_{ij}\in[-0.1,0.1]$ and $[-0.01,0.01]$,
		respectively. The pink, cyan and purple solid lines present the current
		upper experimental limits reported by XENONnT~\cite{XENON:2025vwd},
		LZ~\cite{LZ:2022lsv} and PandaX-4T~\cite{PandaX:2024qfu}
		experiments, respectively.}
	\label{DD_SF}
\end{figure}
The left and right panel in Fig. \ref{DD_SF} show the dependence of
spin-independent (SI) of each DM component $\Omega _{1R}$, $\varphi _{3}$%
-nucleon scattering cross-section on DM masses $m_{\Omega_{1R}}$ and $%
m_{\rho_1}$, within two scenarios of mixed couplings $\lambda_{ij}\in
[-0.1,0.1]$ (blue points) and $\lambda_{ij}\in [-0.01,0.01]$ (green points).
We see that in the left panel, the DM component $\Omega_{1R}$ shows the
satisfied SI-nucleon cross section in both cases of mixed couplings $%
\lambda_{ij}$. This result can be explained because the fermionic DM
candidate $\Omega _{1R}$ interacts with the SM-like Higgs boson due to
mixing $h$ with $H$. The relevant coupling strength is suppressed by $\sin
^{2}\theta _{h}\simeq \frac{v^{2}}{v_{\sigma }^{2}}\ll 1$. Hence, the
SM-like Higgs field portal negligibly contributes to the annihilation
cross-section. Moreover, the $\Omega_{1R}$ scatters with nucleons via only
the SM-like Higgs portal due to their mixing with the new Higgs, $H$, not
interacting with SM quarks and gluons. Therefore, $\Omega_{1R}$ can escape
the current detection of DM, which results the quite low predicted $\sigma^{%
\text{SI}}$. Turning to right panel, the predicted values of DM
component $\varphi_3$ $\sigma^{\text{SI}}_{\varphi_3}$ are shown to be lower
and satisfy measurement limits if $\lambda_{ij}\in [-0.01,0.01]$, compared
to the case $\lambda_{ij}\in [-0.1,0.1]$. This behavior of $\varphi_3$ is
quite consistent as the right panel in Fig .\ref{DD_2S}.

\section{$95$ GeV diphoton excess}

\label{sec:diphoton-excess} 

In this section, we examine the exciting implications of our model for the $%
95~\text{GeV}$ diphoton excess recently reported by the CMS collaboration.
We suggest that this excess in the diphoton final state, observed around $95~%
\text{GeV}$, could be attributed to the real component $\sigma _{R}$ of the
scalar singlet $\sigma$, which we assume has a mass of $95~\text{GeV}$.

The electroweak scalar singlet $\sigma _{R}$ is primarily produced via a
gluon fusion mechanism involving heavy exotic quarks $T_{n}$ and $B_{n}$
(with $n=1,2$) in a triangular loop. Its diphoton decay is mediated by
triangular loops featuring the virtual exchange of vector-like quarks,
charged vector-like leptons $E_{n}$ (for $n=1,2$), and electrically charged
scalars. As a result, the cross-section for the production of the diphoton
scalar resonance at the LHC can be represented as follows: 
\begin{equation}
\sigma _{total}\left( pp\rightarrow \sigma _{R}\rightarrow \gamma \gamma
\right) =\frac{\pi ^{2}}{8}\frac{1}{m_{\sigma _{R}}\Gamma _{\sigma _{R}}}%
\Gamma (\sigma _{R}\rightarrow \gamma \gamma )\frac{1}{s}\int_{\frac{%
m_{\sigma _{R}}^{2}}{s}}^{1}\frac{dx}{x}f_{g}(x)f_{g}\left( \frac{m_{\sigma
_{R}}^{2}}{sx}\right) \Gamma (\sigma _{R}\rightarrow gg),  \label{sigma}
\end{equation}%
where $m_{\sigma _{R}}\simeq 95\text{GeV}$ represents the mass of the
resonance, and $\Gamma _{\sigma _{R}}$ denotes its total decay width. The
function $f_{g}(x)$ is the gluon distribution, and $\sqrt{s}=13$ TeV is the
LHC center of mass energy.

The diphoton excess observed at $95 \text{ GeV} $ can be interpreted as a
scalar resonance with a signal strength described by \cite{CMS:2023yay,
CMS:2024yhz, Biekotter:2023jld}:

\begin{equation}
\mu _{\gamma \gamma }^{\left( \exp \right) }=\frac{\sigma _{\exp }\left(
pp\rightarrow \sigma _{R}\rightarrow \gamma \gamma \right) }{\sigma
_{SM}\left( pp\rightarrow h\rightarrow \gamma \gamma \right) }=0.35\pm 0.12
\,,  \label{mu95GeV}
\end{equation}
where $\sigma _{SM}$ corresponds to the total cross section for a
hypothetical SM Higgs boson at the same mass.

\begin{figure}[h]
\centering
	\includegraphics[width=0.6\linewidth]{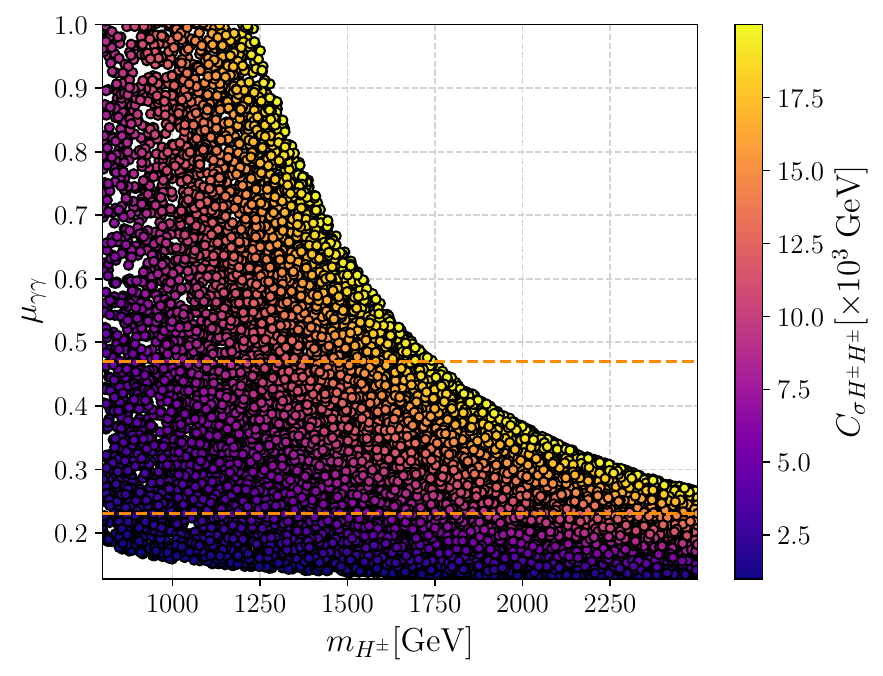}
\caption{Signal strength for a $95$ GeV Diphoton excess as a function of the
charged scalar mass $m_{H^{\pm }}$. The orange horizontal lines
correspond to the $1\protect\sigma $ upper and lower experimental bounds,
respectively.}
\label{fig:97GeVdiphotonexcess}
\end{figure}

The corresponding decay widths of the resonance into photon and gluon pairs
are respectively given by: 
\begin{equation}
\Gamma (\sigma _{R}\rightarrow gg)=\frac{K^{gg}\alpha _{s}^{2}m_{\sigma_R }^{3}%
}{256\pi ^{3}}\left\vert \sum_{n=1}^{2}\frac{y_{T_{n}}}{m_{T_{n}}}%
F(x_{T_{n}})+\sum_{n=1}^{2}\frac{y_{B_{n}}}{m_{B_{n}}}F(x_{B_{n}})\right%
\vert ^{2},  \label{Gammadigluon}
\end{equation}%
\begin{eqnarray}
\Gamma (\sigma _{R}\rightarrow \gamma \gamma ) &=&\frac{\alpha ^{2}m_{\sigma_R
}^{3}}{512\pi ^{3}}\left\vert \sum_{n=1}^{2}\frac{N_{c}Q_{T_{n}}^{2}y_{T_{n}}%
}{m_{T_{n}}}F_{1/2}(\zeta _{T_{n}})+\sum_{n=1}^{2}\frac{%
N_{c}Q_{B_{n}}^{2}y_{B_{n}}}{m_{B_{n}}}F_{1/2}(\zeta _{B_{n}})\right.  \notag
\\
&&\left. +\dsum\limits_{n=1}^{2}\frac{Q_{E_{n}}^{2}y_{E_{n}}}{m_{E_{n}}}%
F_{1/2}(\zeta _{E_{n}})+\frac{C_{\sigma H^{\pm }H^{\mp }}}{\sqrt{2}m_{H^{\pm
}}^{2}}F_{0}(\zeta _{H^{\pm }})\right\vert ^{2}\,,  \label{Gammadiphoton}
\end{eqnarray}%
where $K^{gg}\sim 1.5$ is a QCD loop enhancement factor that accounts for
the higher order QCD corrections, $\zeta _{i}=4M_{i}^{2}/m_{\sigma }^{2}$,
with $M_{i}=m_{T_{n}},m_{B_{n}},m_{E_{n}},m_{H^{\pm }}^{2}$ ($n=1,2$) and
the loop functions $F_{1/2}(\zeta )$ and $F_{0}(\zeta )$ are given by:%
\begin{equation}
F_{1/2}(\zeta )=-2\zeta \left( 1+(1-\zeta )f(\zeta )\right) ,\,\hspace{0.7cm}%
\hspace{0.7cm}F_{0}(\zeta )=\left( 1-\zeta f(\zeta )\right) \zeta ,
\label{F}
\end{equation}%
where 
\begin{equation}
f(\zeta )=\left\{ 
\begin{array}{lcc}
\left\vert \arcsin \sqrt{1/\zeta }\right\vert ^{2} & \text{for} & \zeta \geq
1 \\ 
&  &  \\ 
-\frac{1}{4}\left( \ln \left( \frac{1+\sqrt{1-\zeta }}{1-\sqrt{1-\zeta }}%
\right) -i\pi \right) ^{2} & \text{for} & \zeta <1 \\ 
&  & 
\end{array}%
\right.  \label{f}
\end{equation}%
The Fig. (\ref{fig:97GeVdiphotonexcess}) shows the plot of the ratio $\mu
_{\gamma \gamma }$ as a function of the scalar mass $m_{H^{\pm }}$. In this
plot, the masses of new exotic quarks were estimated using Eq. (\ref%
{eq:mupquarkentries}), which gives $m_{T_{1}}=m_{B_{1}}\simeq 4460$ GeV and $%
m_{T_{2}}=m_{B_{2}}\simeq 3858$ GeV. 
To evaluate the ratio $\mu _{\gamma
\gamma }$, we consider a benchmark scenario where $v_{\sigma }\simeq 5$
TeV, the charged exotic lepton masses are set to be equal to $m_{E_{1}}=m_{E_{2}}\simeq 200 $ GeV and the coupling strength $C_{\sigma H^{\pm }H^{\mp }}$ of the trilinear scalar interaction $H^\pm H\mp \sigma$ takes the values of $2$ to $20$  TeV. These values fit the ratio $\mu _{\gamma \gamma }$ at the $1\sigma $
level for the charged scalar mass in the range $850\leq m_{H^{\pm }}\leq
2500 $ GeV.  As indicated in Fig. (\ref{fig:97GeVdiphotonexcess}), our model
is consistent with the $95$ GeV diphoton excess.  The plot shows that the $\mu
_{\gamma \gamma }$ ratio is sensitive to a variation of the trilinear scalar coupling $C_{\sigma H^{\pm }H^{\mp }}$, then implying, as indicated by Eq. (\ref{Gammadiphoton}), 
that
the contribution of virtual exchange of electrically charged scalars in the triangular loop of the $\sigma$ decay is relevant for $C_{\sigma H^{\pm }H^{\mp }} \sim 5$ to $10$ TeV in a wide range of charged scalar masses $m_{H^{\pm}}$. In addition, increasing the value of the trilinear scalar coupling $C_{\sigma H^{\pm }H^{\mp }}$ yields  an increase that the lower and upper bounds on charged scalar masses $m_{H^{\pm}}$  compatible with the $95$ GeV diphoton excess.  

In our setup, $\sigma_R$ is predominantly a singlet-like state and its
tree-level couplings to SM fermions and gauge bosons arise only through scalar
mixing.
In the parameter region relevant for Fig.~\ref{fig:97GeVdiphotonexcess}, we have
$\tan\theta_h \sim 10^{-1}$--$10^{-2}$. Consequently, both the production rates and the tree-level partial widths into SM final states scale approximately as $\tan^2\theta_h$. 
As a consequence, signal strengths in channels such as $b\bar b$ and
$\tau^+\tau^-$ are generically suppressed to the level
$\mu_{b\bar b},\,\mu_{\tau\tau}\sim\tan^2\theta_h\lesssim 10^{-2}$--$10^{-4}$,
and are not expected to yield observable excesses.
By contrast, the $\gamma\gamma$ mode can be enhanced by the loop contributions
of vector-like charged fermions (and charged scalars), while the total width is
simultaneously reduced due to the suppressed tree-level decays.
This makes the diphoton final state, and potentially $Z\gamma$, the most promising channel for probing $\sigma_R$ near $95~\mathrm{GeV}$ within this model.

\section{Charged lepton flavor violation}

\label{cLFV} The most stringent limits on charged lepton flavor violation
(cLFV) are obtained from the radiative decay $\mu \to e\gamma$. In our
framework, this process arises at the one-loop level and receives
contributions both from light neutrino mixing and from the exchange of heavy
sterile states. The branching ratio can be written as~\cite%
{Langacker:1988up,Lavoura:2003xp,Hue:2017lak} 
\begin{subequations}
\label{br}
\begin{align}
\text{Br}\!\left( l_{i}\rightarrow l_{j}\gamma \right) &= \frac{%
\alpha_{W}^{3}\,s_{W}^{2}\,m_{l_{i}}^{5}}{256\pi ^{2}m_{W}^{4}\Gamma _{i}}
\left| G_{ij}\right| ^{2},  \label{Brmutoegamma1} \\[6pt]
G_{ij} &\simeq \sum_{k=1}^{3} \Big( \big[ (1-RR^{\dagger })U_{\nu }\big]%
^{\ast}\Big)_{ik} \Big( (1-RR^{\dagger })U_{\nu }\Big)_{jk}\, G_{\gamma
}\!\left( \frac{m_{\nu _{k}}^{2}}{m_{W}^{2}}\right)  \notag \\[4pt]
&\quad+\,2\sum_{l=1}^{2}\,(R^{\ast})_{il}\,(R)_{jl}\, G_{\gamma }\!\left( 
\frac{m_{N_{R_{l}}}^{2}}{m_{W}^{2}}\right) ,  \label{Brmutoegamma2} \\[6pt]
G_{\gamma }(x) &= \frac{10-43x+78x^{2}-49x^{3}+18x^{3}\ln x+4x^{4}} {%
12\,(1-x)^{4}} .  \label{Ggamma}
\end{align}

Here $\Gamma _{\mu }=3\times 10^{-19}~\text{GeV}$ is the total muon decay
width, and $U_{\nu}$ denotes the PMNS matrix, since we are working in the
basis where the SM charged-lepton mass matrix is diagonal. The matrix $R$
encodes the mixing between active and heavy states, and is defined as 
\end{subequations}
\begin{equation}
R=\frac{1}{\sqrt{2}}\,m_{D}^{\ast }M^{-1},  \label{eq:Rneutrino}
\end{equation}
with $M$ and $m_D$ the Majorana and Dirac neutrino mass matrices,
respectively.

\begin{figure}[h]
\includegraphics[scale=2]{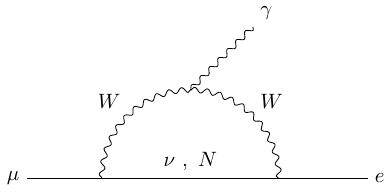}
\caption{One-Loop contribution to $\protect\mu \rightarrow e \protect\gamma$%
. Here, $\protect\nu$ is for light active neutrino and $N$ is for heavy
sterile neutrino both in physical basis.}
\label{fig:muegamma}
\end{figure}

The first term in Eq.~(\ref{br}) corresponds to the modified contribution of
light neutrinos, while the second term arises from loops with heavy sterile
states as it shows is Figure (\ref{fig:muegamma}). Current experimental
limits, most notably from the MEG II experiment \cite{MEGII:2025gzr}, $\text{%
Br}(\mu \to e\gamma) < 1.5\times 10^{-13}$ at $90\%$ C.L., place strong
bounds on the allowed active--sterile mixing parameters contained in $R$.

\begin{figure}[h]
	\includegraphics[width=12cm,height=8.0cm]{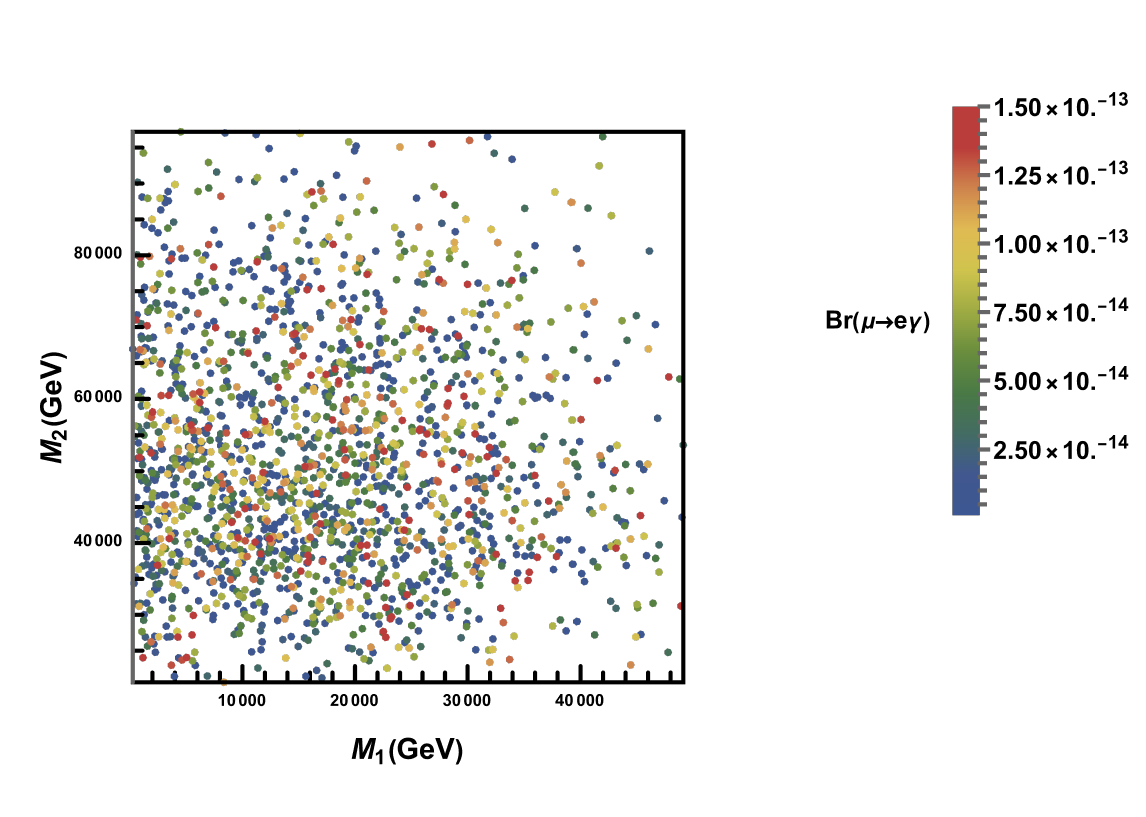}
	\caption{Allowed region in the $M_1$-$M_2$ plane consistent with charged lepton flavor violating constraints.}
	\label{fig:Brmuegamma}
\end{figure}
\begin{figure}[h]
\includegraphics[width=0.6\linewidth]{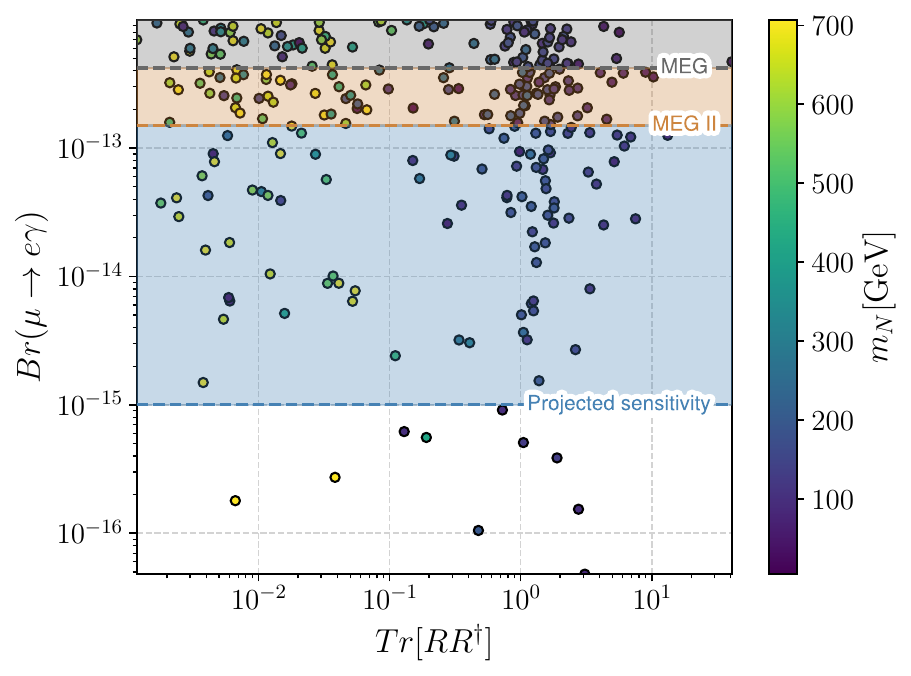}
\caption{Branching ratios of $\mu \rightarrow e \gamma$. The Projected sensitivity dotted line is for hypothetically sensitivity of $10^{-15}$ for future experiments.}
\label{fig:muegamma2}
\end{figure}

To address the concerns regarding quantitative fits, we provide explicit
benchmark points in Table~\ref{tab:benchmark5} for the two-loop inverse
seesaw model. For each benchmark, the Dirac mass matrix $M_D$, the Majorana
mass $M$, and the lepton-number violating parameter $\mu$ are chosen such
that the resulting neutrino mass eigenvalues reproduce the observed
mass-squared differences $\Delta m^2_{21}$ and $|\Delta m^2_{31}|$ within
the current $3\sigma$ experimental ranges~\cite{Esteban:2020cvm}. We stress that the physical PMNS angles $\theta_{12,13,23}$ emerge from the interplay of the neutrino and charged-lepton sectors. Given the significant parametric freedom in this setup, their values cannot be uniquely determined. 
Nevertheless, the corresponding branching ratios for $\mu \to e\gamma$
remain consistent with the latest MEG II bound, demonstrating that our
framework can quantitatively accommodate neutrino oscillation data while
satisfying charged lepton flavor violation constraints. The allowed region in the $M_1$-$M_2$ plane consistent with charged lepton flavor violating constraints is displayed in Figure \ref{fig:Brmuegamma}. This indicates that
the model remains a viable framework for simultaneously explaining neutrino
oscillations and charged lepton flavor violation. In addition, the Figure \ref{fig:muegamma2}, shows the plot of $\mathrm{Br}(\mu \rightarrow e \gamma)$ as a function of $\mathrm{Tr}[R R^{\dagger}]$ from Eq. (\ref{eq:Rneutrino}) and the lightest value of heavy Majorana neutrinos $m_N$. All the points of the plot successfully comply with neutrino oscillation experimental data 
and are consistent with the assumption of diagonal $M$ matrix. The branching ratios of $\mu \rightarrow e \gamma$ are in the reach of actual MEG II sensitivity and can reach the projected sensitvity of $10^{-15}$ for future experiments (Ref. \citep{Bernstein:2013hba} and Ref. \citep{cattaneo2025futureperspectivesmuto}). 

\begin{table}[h!]
\centering
\renewcommand{\arraystretch}{1.3} 
\begin{tabular}{|c|c|c|c|c|}
\hline
Benchmark & $M_D$ (GeV) & $M$ (GeV) & $\mu$ (eV) & Results \\ \hline
1 & $%
\begin{pmatrix}
1.511 & 4.247 \\ 
4.683 & 1.701 \\ 
1.630 & 4.452%
\end{pmatrix}%
$ & $%
\begin{pmatrix}
3742.6 & 0 \\ 
0 & 1729.1%
\end{pmatrix}%
$ & $%
\begin{pmatrix}
455.2 & 3770.7 \\ 
7435.5 & 222.8%
\end{pmatrix}%
$ & 
\begin{tabular}{@{}c}
$Br(\mu \to e \gamma)$=$5.9\times10^{-14}$ \\ 
$\Delta m^2_{21}=7.05\times10^{-5}$ \\ 
$\Delta m^2_{31}=2.48\times10^{-3}$%
\end{tabular}
\\ \hline
2 & $%
\begin{pmatrix}
1.790 & 4.785 \\ 
4.659 & 1.563 \\ 
4.364 & 1.542%
\end{pmatrix}%
$ & $%
\begin{pmatrix}
2402.1 & 0 \\ 
0 & 2920.0%
\end{pmatrix}%
$ & $%
\begin{pmatrix}
418.1 & 3781.0 \\ 
7568.3 & 215.2%
\end{pmatrix}%
$ & 
\begin{tabular}{@{}c}
$Br(\mu \to e \gamma)$=$3.5\times10^{-14}$ \\ 
$\Delta m^2_{21}=7.08\times10^{-5}$ \\ 
$\Delta m^2_{31}=2.38\times10^{-3}$%
\end{tabular}
\\ \hline
3 & $%
\begin{pmatrix}
2.265 & 3.246 \\ 
1.531 & 4.782 \\ 
4.900 & 1.643%
\end{pmatrix}%
$ & $%
\begin{pmatrix}
2656.8 & 0 \\ 
0 & 2341.2%
\end{pmatrix}%
$ & $%
\begin{pmatrix}
330.1 & 3713.0 \\ 
7004.9 & 219.1%
\end{pmatrix}%
$ & 
\begin{tabular}{@{}c}
$Br(\mu \to e \gamma)$=$7.5\times10^{-14}$ \\ 
$\Delta m^2_{21}=6.83\times10^{-5}$ \\ 
$\Delta m^2_{31}=2.53\times10^{-3}$%
\end{tabular}
\\ \hline
4 & $%
\begin{pmatrix}
1.608 & 4.759 \\ 
4.748 & 1.555 \\ 
3.722 & 1.667%
\end{pmatrix}%
$ & $%
\begin{pmatrix}
1989.5 & 0 \\ 
0 & 3209.7%
\end{pmatrix}%
$ & $%
\begin{pmatrix}
403.3 & 3805.5 \\ 
7046.1 & 252.5%
\end{pmatrix}%
$ & 
\begin{tabular}{@{}c}
BR$Br(\mu \to e \gamma)$=$4.7\times10^{-14}$ \\ 
$\Delta m^2_{21}=7.14\times10^{-5}$ \\ 
$\Delta m^2_{31}=2.42\times10^{-3}$%
\end{tabular}
\\ \hline
5 & $%
\begin{pmatrix}
3.880 & 1.707 \\ 
4.293 & 1.501 \\ 
1.579 & 4.664%
\end{pmatrix}%
$ & $%
\begin{pmatrix}
3113.9 & 0 \\ 
0 & 2003.5%
\end{pmatrix}%
$ & $%
\begin{pmatrix}
421.1 & 3807.1 \\ 
7719.3 & 239.9%
\end{pmatrix}%
$ & 
\begin{tabular}{@{}c}
$Br(\mu \to e \gamma)$=$3.6\times10^{-14}$ \\ 
$\Delta m^2_{21}=7.06\times10^{-5}$ \\ 
$\Delta m^2_{31}=2.43\times10^{-3}$%
\end{tabular}
\\ \hline
\end{tabular}%
\caption{Five benchmark points with $M_D$, $M$, $\protect\mu$, branching
ratio $Br(\protect\mu\to e\protect\gamma)$ and neutrino mass-squared
differences.}
\label{tab:benchmark5}
\end{table}

To conclude this section, we discuss the implications of our model for electron-muon conversion. In the low-momentum regime, the dominant contributions to lepton flavor violation (LFV) come from dipole operators, as described by the Effective Lagrangian approach of Ref. \cite{Kuno:1999jp}. This framework establishes relations between the branching ratios of various LFV processes:

\begin{equation}
{\text{Br}}\left( \mu \rightarrow 3e\right) \simeq \frac{1}{160}{\text{Br}}\left( \mu \rightarrow e\gamma \right) ,\quad {\text{CR}}\left( \mu {\text{Ti}}\rightarrow e{\text{Ti}}\right) \simeq \frac{1}{200}{\text{Br}}\left( \mu \rightarrow e\gamma \right) ,\quad {\text{CR}}\left( \mu {\text{Al}}\rightarrow e{\text{Al}}\right) \simeq \frac{1}{350}{\text{Br}}\left( \mu \rightarrow e\gamma \right)
\label{eq:CR-BR}
\end{equation}

The $\mu -e$ conversion ratio, ${\text{CR}}\left( \mu -e\right)$, is defined as~\cite{Lindner:2016bgg}:
\begin{equation}
{\text{CR}}\left( \mu -e\right) =\frac{\Gamma \left( \mu ^{-}+{\text{Nucleus}}\left( A,Z\right) \rightarrow e^{-}+{\text{Nucleus}}\left( A,Z\right) \right) }{\Gamma \left( \mu ^{-}+{\text{Nucleus}}\left( A,Z\right) \rightarrow \nu _{\mu }+{\text{Nucleus}}\left( A,Z-1\right) \right) }
\label{eq:Conversion-Rate}
\end{equation}

Given these relations and our model's prediction for $\mu \rightarrow e\gamma$, the resulting rates for $\mu \rightarrow 3e$ and $\mu -e$ conversion in titanium and aluminum are expected to be $\sim 10^{-15}$. This places them two orders of magnitude below our $\mu \rightarrow e\gamma$ rate and well beneath the current experimental bounds of approximately $10^{-12}$.

\section{Conclusions}

\label{conclusions}

In summary, we have proposed an extension of the inert doublet model that
successfully accommodates some of the unaddressed SM issues. This model is
based on the SM gauge symmetry, supplemented by a spontaneously broken
global $U(1)_X$ symmetry and a preserved $Z_2$ discrete symmetry.

In our model, the first and second families of SM-charged fermions acquire
their masses through a one-loop level radiative seesaw mechanism, while the
third generation of SM-charged fermions obtains their masses at tree level.
The masses of the light-active neutrinos arise from a radiative inverse
seesaw mechanism at two-loop level.

We successfully address the strong CP problem by preserving the CP symmetry
in the SM sector's Lagrangian at the tree level. In this approach, explicit
CP violation occurs in the dark scalar sector. The CP-violating phases are
then transferred to the SM quark sector at the loop level, which is crucial
for providing a viable solution to the strong CP problem. We recover the CP
phase in the weak sector through one-loop level corrections mediated by dark
fields, while the strong CP phase remain vanishing at the three-loop level.

The radiative nature of the seesaw mechanisms is attributed to preserved
discrete symmetries, which are required for ensuring the stability of
fermionic and scalar dark matter candidates. The preserved discrete
symmetries also allow for multi-component dark matter, whose annihilation
processes will enable us to successfully reproduce the measured amount of
dark matter relic abundance while satisfying the current dark matter direct
detection limits.

Additionally, the model complies with 
the constraints related to charged lepton flavor-violating processes, with
the predicted rates for these decays falling within the current experimental
sensitivity.

Finally, we interpret the 95 GeV diphoton excess reported by the CMS
collaboration as a scalar resonance with a mass of 95 GeV, which aligns with
the latest experimental data. 

\section*{\label{acknowledgement}Acknowledgements}
This research was funded by the Vietnam Academy of Science and Technology,
under Grant No. CBCLCA.03/25-27. AECH is supported by ANID-Chile FONDECYT 1241855, 1261103, ANID – Millennium Science Initiative Program $ICN2019\_044$, ANID CCTVal CIA250027 and ICTP through the Associates Programme (2026-2031). N.P. is supported by ANID-Chile Doctorado Nacional año 2022 21221396, and Programa de Incentivo a la Investigaci\'on Cient\'{\i}fica (PIIC) from UTFSM.

\centerline{\bf{REFERENCES}}\vspace{-0.4cm} 
\bibliographystyle{utphys}
\bibliography{Biblio331A4June2022.bib}

\end{document}